\documentclass[aps,prc,superscriptaddress,floats,floatfix,reprint,reprintnumbers]{revtex4}
\usepackage{blindtext}
\usepackage{epsfig}
\usepackage{graphicx}
\usepackage{dcolumn}
\usepackage{bm}
\usepackage{booktabs}
\usepackage{color}
\usepackage{mathrsfs}
\usepackage{ulem}
\usepackage{epstopdf}
\graphicspath{{plots/}}
\usepackage{verbatim}
\usepackage{indentfirst}
\usepackage{float}
\usepackage[colorlinks,
           bookmarks=true,
           linkcolor=blue,
           urlcolor=blue,
            anchorcolor=black,
            citecolor=blue]{hyperref}
\usepackage[bf,raggedright,indentafter]{titlesec}
\usepackage{rotating}
\usepackage{booktabs}
\usepackage{tabularx}
\usepackage{makecell}
\usepackage{hypcap}
\usepackage{subfigure}%
\usepackage{tabularx}

\usepackage{array}

\begin{document}

\hspace{3cm}
\title{\Large Flavor SU(3) analysis  of the charmless semileptonic  $B \to PV\ell^+\nu_\ell$  decays}

\author{Yi Qiao$^1$,~~Jin-Huan Sheng$^2$,~~Yuan-Guo Xu$^3$, and Ru-Min Wang$^{3,\P}$\thanks{ruminwang@sina.com}\\
 {\scriptsize $^1$College of Physics and Electronic Information, Nanchang Normal University, Nanchang, Jiangxi 330032, China}\\
 {\scriptsize $^2$School of Physics and Engineering, Henan University of Science and Technology, Luoyang, Henan, 471000,  China}\\
 {\scriptsize $^3$College of Physics and Communication Electronics, Jiangxi Normal University, Nanchang, Jiangxi, 330022, China}\\
 $^\P${\scriptsize Corresponding author. ~~ruminwang@sina.com}~~
 }

\vspace{3cm}

\begin{abstract}
All  charmless $B \to PV\ell^+\nu_\ell$ ($P$ denotes the light pseudoscalar meson and $V$ denotes the vector meson) decays  have not been experimentally observed to date, but some of them might be measured in the near future.
The  charmless $B \to PV\ell^+\nu_\ell$ decays with  the axial-vector ($A$) resonance states,  the tensor ($T$) resonance states and the excited vector ($V_{E,D}$) resonance states   are studied based on flavor SU(3) analysis in this work, where $V_E$ contains $\{\rho(1450), K^*(1410), \omega(1420), \phi(1680)\}$ with quantum numbers $n^{2S+1}L_J=2^3S_1$ and $V_D$  contains $\{\rho(1700), K^*(1680), \omega(1650), \phi(2170)\}$ with quantum numbers $n^{2S+1}L_J=1^3D_1$.
The  hadronic amplitudes of the $B \to A/T \ell^+\nu_{\ell}$ decays are related by the nonperturbative parameters, and the branching ratio predictions of $B \to A/T/V_{E,D} \ell^+\nu_{\ell}$ are obtained,
and then the branching ratios of the $B\to A(A\to PV)\ell^+\nu_\ell$, $B\to T(T\to PV)\ell^+\nu_\ell$, $B\to V_E(V_E\to PV)\ell^+\nu_\ell$ and $B\to V_D(V_D\to PV)\ell^+\nu_\ell$  are obtained by  the narrow width approximation or further considering the width effects. We find that most branching
ratios of the  $B \to A \ell^+\nu_{\ell}$ decays and  many branching ratios of the $B\to A(A\to PV)\ell^+\nu_\ell$ decays are  on the order of $\mathcal{O}(10^{-4}-10^{-3})$,
 and  all branching ratios with the tensor  resonance states and the excited vector resonance states are small. Our results could be tested in the future BelleII and LHCb experiments.

\end{abstract}
\maketitle

\section{INTRODUCTION}

Semileptonic $B$ decays are very interesting in particle physics due to their relatively clean signals.
Semileptonic $B \to PV\ell^+\nu_\ell$  decays  can receive contributions from both  non-resonant and various resonant contributions. The intermediate resonances in these semileptonic decays provide  a clean environment for exploring  meson spectroscopy. Although no charmless  $B \to PV\ell^+\nu_\ell$ decay  has been measured to date,    near future experiments such as BelleII and LHCb are expected to measure these decays.

Since the weak and strong dynamics can be separated in the semileptonic decays, the theoretical description of the semileptonic decays is relatively simple.  The hadronic
transition form factors contain all the strong dynamics in the initial and final hadrons, making them crucial for testing the theoretical calculations of the involved strong interactions.
However, the calculations of the $B\to PV$ form factors are more complicated than ones for  $B\to V$ or $B\to P$ form factors, and  reliable calculations are currently lacking.
In the absence of reliable calculations,  symmetry analysis plays a vital  role in understanding semileptonic   transitions  by providing relations between hadronic form factors or  hadronic decay amplitudes.
Among these, SU(3) flavor symmetry is one of the symmetries which has
attracted a lot of attention. SU(3) flavor symmetry has been widely used to study hadron decays, including  b-hadron decays \cite{He:1998rq,He:2000ys,Fu:2003fy,Hsiao:2015iiu,He:2015fwa,He:2015fsa,Deshpande:1994ii,Gronau:1994rj,Gronau:1995hm,Shivashankara:2015cta,Zhou:2016jkv,Cheng:2014rfa,Wang:2021uzi,Wang:2020wxn} and
c-hadron decays \cite{Wang:2021uzi,Wang:2020wxn,Grossman:2012ry,Pirtskhalava:2011va,Cheng:2012xb,Savage:1989qr,Savage:1991wu,Altarelli:1975ye,Lu:2016ogy,Geng:2017esc,Geng:2018plk,Geng:2017mxn,Geng:2019bfz,Wang:2017azm,Wang:2019dls,Wang:2017gxe,Muller:2015lua}.

Four-body semileptonic  $D/B \to PP/PV\ell^+\nu_\ell$ decays  have also been explored in previous studies, for instance, in Refs. \cite{Qiao:2024nbq,Wang:2022fbk,Feldmann:2018kqr,Faller:2013dwa,Kim:2017dfr,Ananthanarayan:2005us}.
In this work, we will  focus on analyzing  four-body semileptonic charmless  $B \to A(A \to PV)\ell^+\nu_\ell$, $B \to T(T \to PV)\ell^+\nu_\ell$,  $B\to V_E(V_E\to PV)\ell^+\nu_\ell$ and $B\to V_D(V_D\to PV)\ell^+\nu_\ell$ decays    using the SU(3) flavor symmetry/breaking approach.
The analysis begins by obtaining the form factor relations  for $B \to A/T/V_{E,D} \ell^+ \nu_\ell$ decays through  SU(3) flavor symmetry/breaking. After ignoring  the SU(3) flavor breaking, the branching ratios of the $B \to A/T/V_{E,D} \ell^+ \nu_\ell$ decays are predicted.
Using  these predictions  along with  previous results for $A/T/V_{E,D}\to PV$ decays, the branching ratios for $B\to A(A\to PV)\ell^+\nu_\ell$, $B\to T(T\to PV)\ell^+\nu_\ell$ ,  $B\to V_E(V_E\to PV)\ell^+\nu_\ell$ and $B\to V_D(V_D\to PV)\ell^+\nu_\ell$  decays are obtained by using  the narrow width approximation and considering the width effects.
Due to limited experimental data on  $B\to M\ell^+\nu_\ell$, $M\to PV$  and $B \to M(M \to PV)\ell^+\nu_\ell$ with $M=A,T,V_{E},V_{D}$,  SU(3) flavor  breaking effects are ignored.  The relevant non-perturbative coefficients and  mixing angles are then constrained   by using  existing experimental data. Finally,   the branching ratios that have not yet been measured are predicted by the SU(3) flavor symmetry.

The paper is structured  as follows.  In Section II, three-body semileptonic charmless  $B\to A/T/V_{E,D}\ell^+\nu_{\ell}$ decays are studied.  In Section III, four-body semileptonic charmless  $B\to PV\ell^+\nu_{\ell}$ with the axial-vector resonance states, the tensor  resonance states and the excited vector resonance states are explored. Summary and conclusions are presented in Section IV. Additional details are provided in the Appendixes.

\section{Semileptonic charmless  $B\to M\ell^+\nu_{\ell}$ decays}\label{sec:B2Mlv}
\subsection{Theoretical frame}
The effective Hamiltonian below the electroweak scale can be used to describe the $\bar{b}\to \bar{u} \ell^+\nu_{\ell}$ transitions, which can be written as
\begin{eqnarray}
\mathcal{H}_{eff}(\bar{b} \to \bar{u}\ell^+\nu_\ell)&=&\frac{G_F}{\sqrt{2}}V_{ub}\bar{b}\gamma^\mu(1-\gamma_5)\bar{u}~\bar{\nu_\ell}\gamma_\mu(1-\gamma_5)\ell,
\end{eqnarray}
where $G_F$ is the Fermi constant, and $V_{ub}$ is the Cabibbo-Kobayashi-Maskawa (CKM) matrix element. The decay  amplitude of $B\to M\ell^+\nu_{\ell}$  can be expressed as
\begin{eqnarray}
\mathcal{A}(B\to M\ell^+\nu_\ell)&=&\frac{G_F}{\sqrt{2}}V_{ub}\sum_{m,n}g_{mn} L^{\lambda_\ell\lambda_\nu}_mH^{\lambda_M}_{n},
\end{eqnarray}
with
\begin{eqnarray}
L^{\lambda_\ell\lambda_\nu}_m&=&\epsilon_{\alpha}(m)\bar{\nu_\ell}\gamma_{\alpha}(1-\gamma_5)\ell,\\
H^{\lambda_M}_{n}&=&\epsilon^*_{\beta}(n)\langle{M}(p_M,\varepsilon^*)\bar{u}\gamma^\beta(1-\gamma_5)\bar{b}|B(p_B)\rangle,
\end{eqnarray}
where $\lambda_M=0,\pm1$, $\lambda_\ell=\pm\frac{1}{2}$, and $\lambda_\nu=+\frac{1}{2}$ are the particle helicities, and $\epsilon$ and $\varepsilon^*$ are the polarization vectors of the virtual $W$-boson and $M$-meson, respectively. The leptonic parts $L^{\lambda_\ell\lambda_\nu}_m$ use perturbation theory to calculate, while the hadronic parts $H^{\lambda_M}_{n}$ are encoded in the transition form factors.

The differential branching ratios are \cite{Ivanov:2019nqd}
\begin{eqnarray}
 \frac{d\mathcal{B}(B\to
  M\ell^+\nu_\ell)}{dq^2}=\frac{\tau_{B}G_F^2 |V_{ub}|^2\lambda^{1/2}(q^2-m_\ell^2)^2}{24(2\pi)^3m_B^3q^2}
 {\cal H}_{\rm total}
\end{eqnarray}
with
\begin{eqnarray}
  \label{eq:htot}
&& {\cal H}_{\rm total}\equiv ({\cal H}_U+{\cal
   H}_L)\left(1+\frac{m_\ell^2}{2q^2}\right) +\frac{3m_\ell^2}{2q^2}{\cal H}_S.\\
&& {\cal H}_U=|H^M_+|^2+|H^M_-|^2, \quad {\cal H}_L=|H^M_0|^2, \quad {\cal H}_S=|H^M_t|^2,
\end{eqnarray}
where
$\lambda\equiv
\lambda(m_B^2,m_M^2,q^2)$ with $\lambda(a,b,c)=a^2+b^2+c^2-2ab-2ac-2bc$, $q^2\equiv(p_B-p_M)^2$,    $m_\ell^2\leq q^2\leq(m_B-m_M)^2$, and
$H^M_{\pm,0,t}$  are the hadronic amplitudes.  For  the  $B \to PV\ell^+\nu_\ell$ semileptonic decay, we only consider the axial-vector meson resonant states, the tensor meson resonant states and  the excited vector resonance states.
The hadronic amplitudes  $H^A_{\pm,0,t}$, $H^T_{\pm,0,t}$ and   $H^{V_{E,D}}_{\pm,0,t}$ are listed in Appendix A.
Then the  branching ratios  of the  $B\to M \ell^+\nu_\ell$ decays could be obtained by the relevant form factors, which  depend on the  different methods.

The SU(3) flavor symmetry approach  is independent of the detailed dynamics, and it can provide the hadronic amplitude relationships among various decay modes.
We will use SU(3) flavor symmetry to analyze relative decays in this work.    The SU(3) flavor analysis depends on the SU(3) flavor group, and the relevant meson multiplets of the SU(3) flavor group are listed in Appendix B.

In terms of the SU(3) flavor symmetry, the hadronic amplitudes of the $B\to M\ell^+\nu_{\ell}$ decays can be parametrized as
\begin{eqnarray}  \label{eq:htot}
H(B\to M\ell^+\nu_{\ell})=c^{M}_0B^iM_i^jH_j+c^{M}_1B^aW^i_aM_i^jH_j+c^{M}_2B^iM_i^aW_a^jH_j,
\end{eqnarray}
where $c^{M}_{0}$ ($c^{M}_{1,2}$)  are the SU(3) flavor symmetry (breaking) nonperturbative coefficients.    Please notice that  the  nonperturbative coefficients for  the axial-vector mesons with $J^{PC}=1^{++}$ and for the  axial-vector mesons with $J^{PC}=1^{+-}$ are different, and we will denote them by $c^A_{0,1,2}$ and $c^B_{0,1,2}$, respectively.
 The matrix $W$ is  related to the SU(3) flavor breaking effects, which mainly come from different masses of $u/d$ and $s$ quarks \cite{Xu:2013dta,He:2014xha}.
 The hadronic amplitude  relations for the $B\to M \ell^+\nu_\ell$ decays are summarized in Tab. \ref{Tab:B2MlAmp}.
Amplitude relations between different decay processes can be easily obtained from Tab. \ref{Tab:B2MlAmp}, and the same relationships are also true for the form factors. Since there are not enough experimental data in the $B\to A/T \ell^+\nu_\ell$ decays, we will only consider the SU(3) flavor symmetry contributions, i.e., $C^M=C'^M=c^M_0$.

\begin{table}[h]
\renewcommand\arraystretch{1.0}
\tabcolsep 0.05in
\centering
\caption{SU(3) amplitudes for $B \to M\ell^+\nu_\ell$ decays due to the $\bar{b}\to \bar{u}\ell^+\nu_{\ell}$ transitions. $C^M=c^M_0+c^M_1+c^M_2,~C'^M=c^M_0-2c^M_1+c^M_2$, and $C^M=C'^M$ if ignoring the SU(3) breaking
effects.
}\vspace{0.1cm}
\begin{tabular}{lc|lc}  \hline\hline
Decay modes                           & SU(3) hadronic amplitudes                                                                 &   Decay modes                           & SU(3) hadronic amplitudes\\\hline
$B^+\to a_1(1260)^0\ell^+\nu_\ell$  &$\frac{1}{\sqrt{2}}C^A$                                                                    &$B^+\to a_2(1320)^0\ell^+\nu_\ell$     &$\frac{1}{\sqrt{2}}C^T$\\
$B^+\to f_1(1285)^0\ell^+\nu_\ell$  &$\frac{1}{\sqrt{6}}\Big({\sqrt{2}cos\theta_{3P1}+sin\theta_{3P1}}\Big)C^A$                 &$B^+\to f_2(1270)\ell^+\nu_\ell$       &$\frac{1}{\sqrt{2}}cos\theta_{f_2}C^T$\\
$B^+\to f_1(1420)^0\ell^+\nu_\ell$  &$-\frac{1}{\sqrt{6}}\Big({\sqrt{2}sin\theta_{3P1}-cos\theta_{3P1}}\Big)C^A$                &$B^+\to f_2'(1525)\ell^+\nu_\ell$      &$\frac{1}{\sqrt{2}}sin\theta_{f_2}C^T$\\
$B^0\to a_1(1260)^-\ell^+\nu_\ell$  &$C^A$                                                                                      &$B^0\to a_2(1320)^-\ell^+\nu_\ell$     &$C^T$\\
$B^+\to b_1(1235)^0\ell^+\nu_\ell$  &$\frac{1}{\sqrt{2}}C^B$                                                                    &$B_s^0\to K^*_2(1430)^-\ell^+\nu_\ell$   &$C'^T$\\
$B^+\to h_1(1170)^0\ell^+\nu_\ell$  &$\frac{1}{\sqrt{6}}\Big({\sqrt{2}cos\theta_{1P1}+sin\theta_{1P1}}\Big)C^B$ \\
$B^+\to h_1(1415)^0\ell^+\nu_\ell$  &$-\frac{1}{\sqrt{6}}\Big({\sqrt{2}sin\theta_{1P1}-cos\theta_{1P1}}\Big)C^B$                &$B^+\to \rho(1450)^0\ell^+\nu_\ell/\rho(1700)^0\ell^+\nu_\ell$&$\frac{1}{\sqrt{2}}C^{V_{E/D}}V^*_{ub}$\\
$B^0\to b_1(1235)^-\ell^+\nu_\ell$  &$C^B$                                                                                      &$B^+\to \omega(1420)\ell^+\nu_\ell/\omega(1650)\ell^+\nu_\ell$&$\frac{1}{\sqrt{2}}C^{V_{E/D}}V^*_{ub}$\\
$B_s^0\to K_1(1270)^-\ell^+\nu_\ell$  &$\Big(sin\theta_{K_1}C'^A+cos\theta_{K_1}C'^B\Big)$                                      &$B^0\to \rho(1450)^-\ell^+\nu_\ell/\rho(1700)^-\ell^+\nu_\ell$&$C^{V_{E/D}}V^*_{ub}$\\
$B_s^0\to K_1(1400)^-\ell^+\nu_\ell$  &$\Big(cos\theta_{K_1}C'^A-sin\theta_{K_1}C'^B\Big)$                                      &$B^0_s\to K^*(1410)^-\ell^+\nu_\ell/K^*(1680)^-\ell^+\nu_\ell$&$C'^{V_{E/D}}V^*_{ub}$\\ \hline
\end{tabular}\label{Tab:B2MlAmp}
\end{table}

\subsection{Numerical results}
The theoretical input parameters, such as the lifetimes, the masses, and the
experimental data within the $1\sigma$ error bar from
the Particle  Data  Group (PDG) \cite{PDG2024}
will be used in our numerical analysis.

Until now, none of the $B\to A\ell^+\nu_\ell$ decays have been measured, so we cannot determine the $C^{A,B}$ by the experimental data.  The form factors of the $B\to a_1/b_1$ and $B_s\to K_{1A}/K_{1B}$ transitions  are taken from Ref. \cite{Verma:2011yw}.
For the $B\to f_1(1285)$, $B\to f_1(1420)$, $B\to h_1(1170)$, and $B\to h_1(1415)$ transitions, two cases are considered.
In the ``$a$" case,
we use the    mixing angles $~\theta_{1P_1}\in[5^\circ, 56^\circ]$ and $\theta_{3P_1}\in[56^\circ, 125^\circ]$ defined in Eqs. (\ref{Eq:Mixf1})-(\ref{Eq:Mixh1}) and   the form factors  obtained by using the SU(3) flavor symmetry relationships  from the $B\to a_1/b_1$ form factors.  In the  ``$b$" case, we use the   mixing angles   $\alpha_{f_1}=(69.7\pm8)^\circ$ and $\alpha_{h_1}=(86.7\pm6)^\circ$ defined  in Eqs. (\ref{Eq:alpha1})-(\ref{Eq:alpha2}) and the form factors of the  $B\to f_{1q,s}/h_{1q,s}$ transitions in Ref. \cite{Verma:2011yw}.
The predicted branching ratios of the $B\to A\ell^+\nu_\ell$ decays are listed in Tab. \ref{Tab:B2Arusults}.
One can see that the errors of some results are a little large, the errors of $\mathcal{B}(B\to a_1(1260) \ell^+\nu_{\ell})$ and $\mathcal{B}(B\to b_1(1235) \ell^+\nu_{\ell})$ mainly come from the form factors and the CKM matrix element $V_{ub}$, and the errors of other branching ratios also come from the mixing angles $\theta_{K_1}$, $\theta_{3P_1}$ or $\theta_{1P_1}$.
For the $\mathcal{B}(B^+\to f_1\ell^+\nu_{\ell})$ and  $\mathcal{B}(B^+\to h_1\ell^+\nu_{\ell})$, we listed the results in two cases. We find that every branching ratio is obviously different in two cases  except $\mathcal{B}(B^+\to h_1(1170)\ell^+\nu_{\ell})$.
For the $\mathcal{B}(B_s^0\to K_1^-\ell^+\nu_{\ell})$ decays, their branching ratios are affected by the mixing angle $\theta_{K_1}$,  the results  are obtained by using $~\theta_{K_1}\in[52^\circ, 65^\circ]$  \cite{Wang:2022fbk} and $~\theta_{K_1}=(33\pm4)^\circ$ \cite{Kang:2018jzg}, which are denoted by $^\natural$ and $^\sharp$ in Tab. \ref{Tab:B2Arusults}, respectively. One can see that the predictions of $\mathcal{B}(B_s^0\to K_1(1270)^-\ell^+\nu_{\ell})$ in two cases are similar, nevertheless, the predictions of $\mathcal{B}(B_s^0\to K_1(1400)^-\ell^+\nu_{\ell})$ are slightly different.
\begin{table}[t]
\renewcommand\arraystretch{1.2}
\tabcolsep 0.3in
\centering
\caption{Branching ratios of  the $B\to A\ell^+\nu_\ell$ decays within 1$\sigma$ error (in units of $10^{-4}$); here, $\ell'=e,\mu$.
For $\mathcal{B}(B^+\to f_1/h_1\ell^+\nu_{\ell})$, $^a$denotes the results obtained by using the $\theta_{1P_1}$/$\theta_{3P_1}$  mixing angles and   the form factors  obtained by  the SU(3) flavor symmetry relationships.
$^b$denotes ones obtained by using the    $\alpha_{f_1}$/$\alpha_{h_1}$ mixing angles and the $B\to f_{1q}/h_{1q}$ form factors taken from Ref. \cite{Verma:2011yw}. For $\mathcal{B}(B_s^0\to K_1^-\ell^+\nu_{\ell})$, $^\natural$denotes the results by using  $\theta_{K_1}\in[52^\circ, 65^\circ]$ \cite{Wang:2022fbk}, and $^\sharp$denotes ones by using $~\theta_{K_1}=(33\pm4)^\circ$  \cite{Kang:2018jzg}. The  $a,b,\natural,\sharp$ signs are same in Tables   \ref{Tab:B2PVlv1} and \ref{Tab:B2PVlv2}.
}\vspace{0.1cm}
\begin{tabular}{lcc}  \hline\hline
Branching ratios                &      Predictions with $\ell=\ell'$               &      Predictions with $\ell=\tau$  \\
$\mathcal{B}(B^+\to a_1(1260)^0 \ell^+\nu_{\ell})$              &      $16.33\pm4.01$               &      $6.91\pm1.87$\\
$\mathcal{B}(B^+\to b_1(1235)^0 \ell^+\nu_{\ell})$              &      $28.16\pm10.01$              &      $13.26\pm4.42$\\
$\mathcal{B}(B^+\to f_1(1285)^0\ell^+\nu_{\ell})$               &      $^{[8.32\times10^{-4},15.67]^a}_{13.39\pm3.99^b}$               &      $^{[3.35\times10^{-4},6.43]^a}_{5.44\pm1.69^b}$\\
$\mathcal{B}(B^+\to f_1(1420)^0\ell^+\nu_{\ell})$               &      $^{9.60\pm7.99^a}_{2.24\pm1.69^b}$                &      $^{3.87\pm3.23^a}_{0.90\pm0.67^b}$\\

$\mathcal{B}(B^+\to h_1(1170)^0\ell^+\nu_{\ell})$               &      $^{30.87\pm11.65^a}_{32.82\pm12.32^b}$              &      $^{14.79\pm4.95^a}_{15.11\pm5.97^b}$\\
$\mathcal{B}(B^+\to h_1(1415)^0 \ell^+\nu_{\ell})$              &      $^{[5.37\times10^{-8},5.82]^a}_{[2.82\times10^{-7},0.56]^b}$   &      $^{[2.36\times10^{-8},2.45]^a}_{[1.21\times10^{-7},0.24]^b}$\\
$\mathcal{B}(B^0\to a_1(1260)^-\ell^+\nu_{\ell})$               &      $30.31\pm7.39$               &      $12.85\pm3.47$\\
$\mathcal{B}(B^0\to b_1(1235)^-\ell^+\nu_{\ell})$               &      $52.21\pm18.57$              &      $24.63\pm8.23$\\
$\mathcal{B}(B_s^0\to K_1(1270)^-\ell^+\nu_{\ell})$               &      $^{72.42\pm19.60^\natural}_{74.43\pm19.53^\sharp}$              &      $^{31.24\pm9.18^\natural}_{33.19\pm9.41^\sharp}$\\
$\mathcal{B}(B_s^0\to K_1(1400)^-\ell^+\nu_{\ell})$               &      $^{5.73\pm3.82^\natural}_{3.03\pm2.16^\sharp}$                &      $^{2.89\pm1.96^\natural}_{1.17\pm0.89^\sharp}$\\
 \hline
\end{tabular}\label{Tab:B2Arusults}
%
\renewcommand\arraystretch{1.1}
\tabcolsep 0.2in
\centering
\caption{Branching ratios of  $B\to T\ell^+\nu_{\ell}$ decays within 1$\sigma$ error (in units of $10^{-5}$). $^\dag$denotes $\mathcal{B}(B^+\to f_2(1270)\ell^+\nu_\ell)$ are obtained by
$\mathcal{B}(B^+\to \pi^+\pi^-\ell'^+\nu_{\ell'})^{Exp.}_{f_2(1270)}/\mathcal{B}(f_2(1270)\to\pi^+\pi^- )^{Exp.}$  from Ref. \cite{PDG2024}.}\vspace{0.1cm}
\begin{tabular}{lccc}  \hline\hline
Branching ratios                                          &   Our predictions$ (\ell=\ell')$      &   Our predictions $(\ell=\tau )$\\\hline
$\mathcal{B}(B^+\to a_2(1320)^0\ell^+\nu_\ell)$         &  $2.92\pm1.51$                        &  $0.93\pm0.58$\\
$\mathcal{B}(B^+\to f_2(1270)\ell^+\nu_\ell)$           &  $3.21\pm1.65^\dag$                   &  $1.06\pm0.66$\\
$\mathcal{B}(B^+\to f'_2(1525)\ell^+\nu_\ell)$          &  $(4.56\pm3.03)\times10^{-2}$         &  $(1.27\pm0.91)\times10^{-2}$\\
$\mathcal{B}(B^0\to a_2(1320)^-\ell^+\nu_\ell)$         &  $5.41\pm2.79$                        &  $1.72\pm1.07$\\
$\mathcal{B}(B^0_s\to K_2(1430)^-\ell^+\nu_\ell)$         &  $4.72\pm2.48$                        &  $1.50\pm0.94$\\\hline
\end{tabular}\label{Tab:BrB2Tlv}
\renewcommand\arraystretch{1.1}
\tabcolsep 0.2in
\centering
\caption{Branching ratios of  $B\to V_{E/D}\ell^+\nu_{\ell}$ decays within 1$\sigma$ error (in units of $10^{-5}$).
$^a$denotes the form factors calculated by the large energy effective theory, and $^b$denotes the form factors calculated by  the light-front quark models. }\vspace{0.1cm}
\begin{tabular}{lcc}  \hline\hline
Branching ratios                                          &   Our predictions$ (\ell=\ell')$    &   Our predictions $(\ell=\tau )$            \\\hline
$\mathcal{B}(B^+\to \rho(1450)^0\ell^+\nu_\ell)$          &  $10.36\pm1.52^a$,~  $8.14\pm1.15^b$                 &  $4.43\pm0.71^a$,~  $3.18\pm0.48^b$                         \\
$\mathcal{B}(B^+\to \omega(1420)\ell^+\nu_\ell)$          &  $11.34\pm2.20^a$,~  $9.00\pm1.77^b$                 &  $5.02\pm1.14^a$,~  $3.64\pm0.82^b$                         \\
$\mathcal{B}(B^0\to \rho(1450)^-\ell^+\nu_\ell)$          &  $19.26\pm2.87^a$,~  $15.14\pm2.15^b$                &  $8.23\pm1.34^a$,~  $5.91\pm0.91^b$                         \\
$\mathcal{B}(B^0_s\to K^*(1410)^-\ell^+\nu_\ell)$         &  $24.14\pm3.17^a$,~  $23.69\pm3.19^b$                &  $11.21\pm1.55^a$,~  $10.16\pm1.44^b$                          \\\hline
$\mathcal{B}(B^+\to \rho(1700)^0\ell^+\nu_\ell)$          &  $5.25\pm0.73^a$                                     &  $1.87\pm0.29^a$                             \\
$\mathcal{B}(B^+\to \omega(1650)\ell^+\nu_\ell)$          &  $5.66\pm0.85^a$                                     &  $2.10\pm0.35^a$                           \\
$\mathcal{B}(B^0\to \rho(1700)^-\ell^+\nu_\ell)$          &  $9.77\pm1.35^a$                                     &  $3.47\pm0.53^a$                              \\
$\mathcal{B}(B^0_s\to K^*(1680)^-\ell^+\nu_\ell)$         &  $11.35\pm1.55^a$                                      &  $4.33\pm0.65^a$                             \\\hline
\end{tabular}\label{Tab:BrB2VEDlv}\end{table}

As for the $B \to T \ell^+\nu_\ell$ decays,   no decay mode has been measured. The data of $\mathcal{B}(B^+\to f_2(1270)\ell'^+\nu_{\ell'})$ may be obtained  by
$\mathcal{B}(B^+\to f_2(1270)^0\ell'^+\nu_{\ell'})=\frac{\mathcal{B}(B^+\to\pi^+\pi^-\ell'^+\nu_{\ell'} )_{f_2(1270)^0}^{Exp.}}{\mathcal{B}(f_2(1270)^0\to \pi^+\pi^-)^{Exp.}}$ from Ref. \cite{PDG2024}, as follows:
\begin{eqnarray}
\mathcal{B}(B^+ \to f_2(1270)^0 \ell'^+\nu_{\ell'})^{Exp.}=(3.21\pm1.65)\times10^{-5}. \label{eq:BB2f1270}
\end{eqnarray}
The relations of the $B \to T \ell^+\nu_\ell$ decays  in Tab. \ref{Tab:B2MlAmp} will be used for the form factors  $A^T_1(0)$.  Other form factors $F_i(0)=V^T(0),A^T_0(0),A^T_2(0)$ can be expressed as $r_i\times A^T_1(0)$,  and the values of the ratios $r_i=\frac{F_i(0)}{A^T_1(0)}$ are taken from Ref. \cite{Chen:2021ywv}. The form factor  $A^T_1(0)$ is determined by the experimental data given in Eq. (\ref{eq:BB2f1270}), and we obtain $A^T_1(0)=0.14\pm0.06$ ( it only include part, such as $C^T$ but not $\frac{1}{\sqrt{2}}cos\theta_{f_2}$), which is slightly smaller than $0.19^{+0.04}_{-0.03}$ in Ref. \cite{Chen:2021ywv}.
 Then the branching ratios of the $B\to T\ell^+\nu_{\ell}$  decays can be predicted in terms of the constrained $A^T_1(0)$, and they are listed in  Tab. \ref{Tab:BrB2Tlv}.

For the $B \to V_{E,D}\ell^+\nu_\ell$ decays,   there is still no  measured mode.   The form factors of the $B\to K^*(1410)$ and $B\to K^*(1680)$ transitions are taken from Ref. \cite{Hatanaka:2010fpr},
which are calculated by the large energy effective theory, and
other form factors of $B\to V_{E,D}$ are related with them by the SU(3) flavor symmetry given in Tab. \ref{Tab:B2MlAmp}.
Then the branching ratios of the $B\to V_{E,D}\ell^+\nu_{\ell}$  decays  are obtained and listed in  Tab. \ref{Tab:BrB2VEDlv}.
In addition, the form factors of $B\to V_E$ transitions are also calculated by the light-front quark models \cite{CQPrepare}, and they are listed in Tab. \ref{Tab:FF} of Appendix A.
The branching ratios of the $B\to V_E \ell^+\nu_{\ell}$ decays are also predicted by using the form factors within  the light-front quark models, and they are given in  Tab. \ref{Tab:BrB2VEDlv}, too.
One can see that branching ratios obtained by the light-front quark models are slightly smaller than ones obtained by the large energy effective theory. In the following analysis, the results from  the large energy effective theory
will be used.

\section{Charmless  decays $B\to PV\ell^+\nu_{\ell}$  with the resonances}
\subsection{ $\mathcal{B}(B\to PV\ell^+\nu_{\ell})$ with the narrow width approximation}
For the  $B\to PV\ell^+\nu_{\ell}$ decays with the resonances,
if the decay widths of the resonance  states are very narrow, the resonant branching ratios can be written as
\begin{equation}
\mathcal{B}(B \rightarrow R \ell^+\nu_\ell, R \rightarrow PV) = \mathcal{B}(B \rightarrow R \ell^+\nu_\ell) \times  \mathcal{B}(R \rightarrow PV),\label{Eq:Br4BD}
\end{equation}
where $R$ denotes the  resonances, and $R=A/T$ in this work.   This relation is also a good approximation for wider resonances \cite{Kim:2017dfr}.
The branching ratios of the two-body non-leptonic decays $\mathcal{B}(A\rightarrow PV)$  and $\mathcal{B}(T\rightarrow PV)$  have been studied in our previous work \cite{Wang:2022fbk}.  Using the expressions  of $\mathcal{B}(B\to A/T\ell^+\nu_\ell)$ given in Sec. \ref{sec:B2Mlv}  and    $\mathcal{B}(A/T\rightarrow PV)$ given in Ref. \cite{Wang:2022fbk}, we may  predict the branching ratios of the $B\to A(A\to PV)\ell^+\nu_\ell$ and $B\to T(T\to PV)\ell^+\nu_\ell$ decays, which are listed in the second and third columns of Tabs. \ref{Tab:B2PVlv1} and \ref{Tab:B2PVlv2}, respectively, named as the results in $S_1$ case.
\begin{table}[b]
\renewcommand\arraystretch{0.95}
\tabcolsep 0.01in
\centering
\caption{SU(3) branching ratios of  $B\to PV\ell'^+\nu_{\ell'}$ decays within 1$\sigma$ error  (in units of $10^{-4}$). $\mathcal{B}_{[A/T]}$  denote the branching ratio with the axial-vector/tensor  meson resonant contributions. In the $S_1$ case, the results are obtained by using the narrow width approximation in Eq. (\ref{Eq:Br4BD}).  In the $S_2$ case, the width effects are considered as given in Eq.(\ref{Eq:RDA4}).  The definitions of $\mathcal{B}_{A/T}$ and $S_{1,2}$   cases are same  in Table  \ref{Tab:B2PVlv2}.}\vspace{0.08cm}
{\footnotesize
\begin{tabular}{l|ll|ll}  \hline\hline
Decay modes & $\mathcal{B}_{[A]}$ in $S_1$ case & $\mathcal{B}_{[T]}$ in $S_1$ case & $\mathcal{B}_{[A]}$ in $S_2$ case  & $\mathcal{B}_{[T]}$ in $S_2$ case\\\hline
$B^+\to \rho^-\pi^+\ell'^+\nu_{\ell'}$                   &$6.58\pm2.71_{[a_1(1260)^0]}$                                                &$0.10\pm0.06_{[a_2(1320)^0]}$                     &$3.68\pm1.42_{[a_1(1260)^0]}$                    &$0.090\pm0.048_{[a_2(1320)^0]}$       \\
or                                                       &$^{6.93\pm3.85^a}_{7.53\pm3.10^b}$$_{[h_1(1170)^0]}$                         &                                                  &$^{4.27\pm2.45^a}_{4.59\pm1.62^b}$$_{[h_1(1170)^0]}$                    &                                    \\
$B^+\to \rho^+\pi^-\ell'^+\nu_{\ell'}$                   &$^{0.64\pm0.64^a}_{0.045\pm0.045^b}$$_{[h_1(1415)^0]}$                       &                                                  &$^{0.54\pm0.54^a}_{0.029\pm0.029^b}$$_{[h_1(1415)^0]}$                  &                                    \\\hline
$B^+\to \rho^0\pi^0\ell'^+\nu_{\ell'}$                   &$^{6.96\pm3.87^a}_{7.57\pm3.12^b}$$_{[h_1(1170)^0]}$                         &                                                  &$^{4.35\pm2.50^a}_{4.66\pm1.69^b}$$_{[h_1(1170)^0]}$                    &                                    \\
                                                         &$^{0.64\pm0.64^a}_{0.045\pm0.045^b}$$_{[h_1(1415)^0]}$                       &                                                  &$^{0.54\pm0.54^a}_{0.029\pm0.029^b}$$_{[h_1(1415)^0]}$                  &                                    \\\hline
$B^+\to \omega\eta\ell'^+\nu_{\ell'}$                    &$^{0.66\pm0.35^a}_{0.68\pm0.26^b}$$_{[h_1(1170)^0]}$                         &                                                  &$^{0.24\pm0.12^a}_{0.25\pm0.09^b}$$_{[h_1(1170)^0]}$                    &                                    \\
                                                         &$^{0.150\pm0.150^a}_{0.012\pm0.012^b}$$_{[h_1(1415)^0]}$                     &                                                  &$^{0.10\pm0.10^a}_{0.0067\pm0.0067^b}$$_{[h_1(1415)^0]}$                  &                                    \\\hline
$B^+\to \phi\eta\ell'^+\nu_{\ell'}$                      &$^{0.022\pm0.022^a}_{0.018\pm0.018^b}$$_{[h_1(1170)^0]}$                     &                                                  &$^{0.0053\pm0.0053^a}_{0.0055\pm0.0055^b}$$_{[h_1(1170)^0]}$                  &                                    \\
                                                         &$^{0.0225\pm0.0225^a}_{0.0032\pm0.0032^b}$$_{[h_1(1415)^0]}$                 &                                                  &$^{0.0050\pm0.0050^a}_{0.00061\pm0.00061^b}$$_{[h_1(1415)^0]}$                &                                    \\\hline
$B^+\to \omega\pi^0\ell'^+\nu_{\ell'}$                   &$18.50\pm7.32_{[b_1(1235)^0]}$                                               &                                                  &$15.59\pm5.70_{[b_1(1235)^0]}$                    &                                    \\\hline
$B^+\to \rho^0\eta\ell'^+\nu_{\ell'}$                    &$1.44\pm0.61_{[b_1(1235)^0]}$                                                &                                                  &$0.41\pm0.17_{[b_1(1235)^0]}$                    &                                    \\\hline
$B^+\to K^{*+}K^-\ell'^+\nu_{\ell'}$                     &$0.32\pm0.14_{[a_1(1260)^0]}$                                                &$0.00016\pm0.00009_{[a_2(1320)^0]}$               &$0.15\pm0.06_{[a_1(1260)^0]}$                    &$0.000079\pm0.000046_{[a_2(1320)^0]}$          \\
or                                                       &$0.12\pm0.05_{[b_1(1235)^0]}$                                                &$0.00025\pm0.00015_{[f_2(1270)^0]}$               &$0.059\pm0.021_{[b_1(1235)^0]}$                   &$0.000095\pm0.000058_{[f_2(1270)^0]}$          \\
$B^+\to K^{*-}K^+\ell'^+\nu_{\ell'}$                     &$^{0.045\pm0.045^a}_{0.08\pm0.04^b}$$_{[f_1(1285)^0]}$                       &$0.00028\pm0.00020_{[f_2'(1525)^0]}$              &$_{0^{a,b}}$$_{[f_1(1285)^0]}$                  &$0.00022\pm0.00015_{[f_2'(1525)^0]}$         \\
                                                         &$^{2.20\pm2.20^a}_{0.47\pm0.47^b}$$_{[f_1(1420)^0]}$                         &                                                  &$^{1.34\pm1.34^a}_{0.27\pm0.27^b}$$_{[f_1(1420)^0]}$                    &                                             \\
                                                         &$^{0.27\pm0.20^a}_{0.29\pm0.21^b}$$_{[h_1(1170)^0]}$                         &                                                  &$^{0.081\pm0.057^a}_{0.094\pm0.062^b}$$_{[h_1(1170)^0]}$                    &                                             \\
                                                         &$^{0.127\pm0.127^a}_{0.037\pm0.037^b}$$_{[h_1(1415)^0]}$                     &                                                  &$^{0.096\pm0.096^a}_{0.024\pm0.024^b}$$_{[h_1(1415)^0]}$                  &                                             \\\hline
$B^+\to K^{*0}\overline{K}^0\ell'^+\nu_{\ell'}$          &$0.31\pm0.13_{[a_1(1260)^0]}$                                                &$0.00013\pm0.00007_{[a_2(1320)^0]}$               &$0.14\pm0.06_{[a_1(1260)^0]}$                    &$0.000064\pm0.000037_{[a_2(1320)^0]}$          \\
or                                                       &$0.11\pm0.04_{[b_1(1235)^0]}$                                                &$0.00022\pm0.00013_{[f_2(1270)^0]}$               &$0.051\pm0.018_{[b_1(1235)^0]}$                    &$0.000085\pm0.000052_{[f_2(1270)^0]}$          \\
$B^+\to \overline{K}^{*0}K^0\ell'^+\nu_{\ell'}$          &$^{0.012\pm0.012^a}_{0.02\pm0.01^b}$$_{[f_1(1285)^0]}$                       &$0.00025\pm0.00018_{[f_2'(1525)^0]}$              &$_{0^{a,b}}$$_{[f_1(1285)^0]}$                  &$0.00019\pm0.00014_{[f_2'(1525)^0]}$         \\
                                                         &$^{1.96\pm1.96^a}_{0.42\pm0.42^b}$$_{[f_1(1420)^0]}$                         &                                                  &$^{1.19\pm1.19^a}_{0.24\pm0.24^b}$$_{[f_1(1420)^0]}$                    &                                             \\
                                                         &$^{0.26\pm0.19^a}_{0.28\pm0.20^b}$$_{[h_1(1170)^0]}$                         &                                                  &$^{0.075\pm0.053^a}_{0.089\pm0.059^b}$$_{[h_1(1170)^0]}$                  &                                             \\
                                                         &$^{0.109\pm0.109^a}_{0.032\pm0.032^b}$$_{[h_1(1415)^0]}$                     &                                                  &$^{0.085\pm0.085^a}_{0.021\pm0.021^b}$$_{[h_1(1415)^0]}$                  &                                             \\\hline
$B^0\to \rho^0\pi^-\ell'^+\nu_{\ell'}$                   &$12.19\pm5.02_{[a_1(1260)^-]}$                                               &$0.19\pm0.10_{[a_2(1320)^-]}$                     &$6.84\pm2.63_{[a_1(1260)^-]}$                    &$0.17\pm0.09_{[a_2(1320)^-]}$                   \\\hline
$B^0\to \rho^-\pi^0\ell'^+\nu_{\ell'}$                   &$12.25\pm5.06_{[a_1(1260)^-]}$                                               &$0.19\pm0.10_{[a_2(1320)^-]}$                     &$6.89\pm2.64_{[a_1(1260)^-]}$                    &$0.17\pm0.09_{[a_2(1320)^-]}$                   \\\hline
$B^0\to \omega\pi^-\ell'^+\nu_{\ell'}$                   &$34.16\pm13.48_{[b_1(1235)^-]}$                                              &                                                  &$6.37\pm2.26_{[b_1(1235)^-]}$                   &                                                \\\hline
$B^0\to \rho^-\eta\ell'^+\nu_{\ell'}$                    &$2.69\pm1.14_{[b_1(1235)^-]}$                                                &                                                  &$0.52\pm0.22_{[b_1(1235)^-]}$                    &                                                \\\hline
$B^0\to K^{*0}K^-\ell'^+\nu_{\ell'}$                     &$1.17\pm0.50_{[a_1(1260)^-]}$                                                &$0.00053\pm0.00030_{[a_2(1320)^-]}$               &$0.54\pm0.22_{[a_1(1260)^-]}$                    &$0.00027\pm0.00015_{[a_2(1320)^-]}$              \\
                                                         &$0.42\pm0.17_{[b_1(1235)^-]}$                                                &                                                  &$0.21\pm0.08_{[b_1(1235)^-]}$                    &                                                 \\\hline
$B^0\to K^{*-}K^0\ell'^+\nu_{\ell'}$                     &$1.18\pm0.51_{[a_1(1260)^-]}$                                                &$0.00056\pm0.00032_{[a_2(1320)^-]}$               &$0.54\pm0.22_{[a_1(1260)^-]}$                    &$0.00026\pm0.00015_{[a_2(1320)^-]}$               \\
                                                         &$0.43\pm0.17_{[b_1(1235)^-]}$                                                &                                                  &$0.21\pm0.09_{[b_1(1235)^-]}$                    &                                                  \\\hline
$B_s^0\to K^{*-}\pi^0\ell'^+\nu_{\ell'}$                 &$^{5.39\pm3.23^\natural}_{5.60\pm3.40^\sharp}$$_{[K_1(1270)^-]}$                         &$0.040\pm0.022_{[K_2^*(1430)^-]}$                   &$^{4.48\pm2.50^\natural}_{4.55\pm2.50^\sharp}$$_{[K_1(1270)^-]}$                    &$0.035\pm0.020_{[K_2^*(1430)^-]}$                   \\
                                                         &$^{1.73\pm1.19^\natural}_{0.95\pm0.69^\sharp}$$_{[K_1(1400)^-]}$                         &                                                  &$^{1.20\pm0.83^\natural}_{0.82\pm0.58^\sharp}$$_{[K_1(1400)^-]}$                    &                                                    \\\hline
$B_s^0\to \overline{K}^{*0}\pi^-\ell'^+\nu_{\ell'}$      &$^{10.58\pm6.34^\natural}_{10.97\pm6.66^\sharp}$$_{[K_1(1270)^-]}$                       &$0.076\pm0.043_{[K_2^*(1430)^-]}$                   &$^{8.76\pm4.87^\natural}_{8.89\pm4.89^\sharp}$$_{[K_1(1270)^-]}$                    &$0.067\pm0.038_{[K_2^*(1430)^-]}$                    \\
                                                         &$^{3.42\pm2.35^\natural}_{1.89\pm1.36^\sharp}$$_{[K_1(1400)^-]}$                         &                                                  &$^{2.36\pm1.64^\natural}_{1.61\pm1.15^\sharp}$$_{[K_1(1400)^-]}$                    &                                                     \\\hline
$B_s^0\to \rho^0K^-\ell'^+\nu_{\ell'}$                   &$^{12.39\pm3.49^\natural}_{12.64\pm3.38^\sharp}$$_{[K_1(1270)^-]}$                        &$0.014\pm0.008_{[K_2^*(1430)^-]}$                 &$^{5.73\pm1.47^\natural}_{5.83\pm1.33^\sharp}$$_{[K_1(1270)^-]}$                    &$0.015\pm0.009_{[K_2^*(1430)^-]}$                    \\
                                                         &$^{0.066\pm0.066^\natural}_{0.022\pm0.022^\sharp}$$_{[K_1(1400)^-]}$                     &                                                  &$^{0.037\pm0.037^\natural}_{0.016\pm0.016^\sharp}$$_{[K_1(1400)^-]}$                    &                                                     \\\hline
$B_s^0\to \rho^-\overline{K}^0\ell'^+\nu_{\ell'}$        &$^{23.97\pm6.63^\natural}_{24.51\pm6.56^\sharp}$$_{[K_1(1270)^-]}$                       &$0.027\pm0.016_{[K_2^*(1430)^-]}$                 &$^{10.62\pm2.73^\natural}_{10.88\pm2.49^\sharp}$$_{[K_1(1270)^-]}$                    &$0.028\pm0.017_{[K_2^*(1430)^-]}$                    \\
                                                         &$^{0.13\pm0.13^\natural}_{0.044\pm0.044^\sharp}$$_{[K_1(1400)^-]}$                       &                                                  &$^{0.072\pm0.072^\natural}_{0.032\pm0.032^\sharp}$$_{[K_1(1400)^-]}$                    &                                                     \\\hline
$B_s^0\to \omega K^-\ell'^+\nu_{\ell'}$                  &$^{6.88\pm1.92^\natural}_{7.01\pm1.87^\sharp}$$_{[K_1(1270)^-]}$                         &$0.015\pm0.011_{[K_2^*(1430)^-]}$                 &$^{4.94\pm1.30^\natural}_{5.08\pm1.18^\sharp}$$_{[K_1(1270)^-]}$                    &$0.016\pm0.011_{[K_2^*(1430)^-]}$                    \\
                                                         &$^{0.064\pm0.064^\natural}_{0.021\pm0.021^\sharp}$$_{[K_1(1400)^-]}$                     &                                                  &$^{0.035\pm0.035^\natural}_{0.016\pm0.016^\sharp}$$_{[K_1(1400)^-]}$                    &                                                     \\\hline
$B_s^0\to K^{*-}\eta\ell'^+\nu_{\ell'}$                  &$^{0.37\pm0.19^\natural}_{0.38\pm0.19^\sharp}$$_{[K_1(1270)^-]}$                         &$0.0028\pm0.0017_{[K_2^*(1430)^-]}$               &$^{0.15\pm0.08^\natural}_{0.14\pm0.08^\sharp}$$_{[K_1(1270)^-]}$                    &$0.0014\pm0.0009_{[K_2^*(1430)^-]}$                  \\
                                                         &$^{0.15\pm0.10^\natural}_{0.11\pm0.10^\sharp}$$_{[K_1(1400)^-]}$                         &                                                  &$^{0.078\pm0.052^\natural}_{0.080\pm0.068^\sharp}$$_{[K_1(1400)^-]}$                    &                                                     \\\hline
$B_s^0\to \phi K^-\ell'^+\nu_{\ell'}$                    & $^{0.22\pm0.15^\natural}_{0.12\pm0.09^\sharp}$$_{[K_1(1400)^-]}$                        &$0.00019\pm0.00012_{[K_2^*(1430)^-]}$             & $^{0.11\pm0.08^\natural}_{0.082\pm0.059^\sharp}$$_{[K_1(1400)^-]}$                   &$0.00012\pm0.00007_{[K_2^*(1430)^-]}$              \\\hline
\end{tabular}}\label{Tab:B2PVlv1}
\end{table}
\begin{table}
\renewcommand\arraystretch{0.95}
\tabcolsep 0.01in
\centering
\caption{SU(3) branching ratios of  $B^+\to PV\tau^+\nu_\tau$ decays within 1$\sigma$ error  (in units of $10^{-4}$). }\vspace{0.08cm}
{\footnotesize
\begin{tabular}{l|ll|ll}  \hline\hline
Decay modes                                            & $\mathcal{B}_{[A]}$ in $S_1$ case                                 & $\mathcal{B}_{[T]}$ in $S_1$ case            & $\mathcal{B}_{[A]}$ in $S_2$ case            & $\mathcal{B}_{[T]}$ in $S_2$ case\\\hline
$B^+\to \rho^-\pi^+\tau^+\nu_\tau$                     &$2.79\pm1.19_{[a_1(1260)^0]}$                                      &$0.033\pm0.021_{[a_2(1320)^0]}$                 &$1.41\pm0.56_{[a_1(1260)^0]}$                 &$0.028\pm0.018_{[a_2(1320)^0]}$       \\
or                                                     &$^{3.22\pm1.67^a}_{3.56\pm1.30^b}$$_{[h_1(1170)^0]}$               &                                              &$^{1.98\pm1.07^a}_{2.08\pm0.67^b}$$_{[h_1(1170)^0]}$                 &                                    \\
$B^+\to \rho^+\pi^-\tau^+\nu_\tau$                     &$^{0.27\pm0.27^a}_{0.019\pm0.019^b}$$_{[h_1(1415)^0]}$             &                                              &$^{0.23\pm0.23^a}_{0.013\pm0.013^b}$$_{[h_1(1415)^0]}$               &                                    \\\hline
$B^+\to \rho^0\pi^0\tau^+\nu_\tau$                     &$^{3.24\pm1.67^a}_{3.58\pm1.31^b}$$_{[h_1(1170)^0]}$               &                                              &$^{2.00\pm1.09^a}_{2.08\pm0.67^b}$$_{[h_1(1170)^0]}$                 &                                    \\
                                                       &$^{0.27\pm0.27^a}_{0.019\pm0.019^b}$$_{[h_1(1415)^0]}$             &                                              &$^{0.23\pm0.23^a}_{0.013\pm0.013^b}$$_{[h_1(1415)^0]}$               &                                    \\\hline
$B^+\to \omega\eta\tau^+\nu_\tau$                      &$^{0.30\pm0.15^a}_{0.32\pm0.12^b}$$_{[h_1(1170)^0]}$               &                                              &$^{0.094\pm0.049^a}_{0.10\pm0.04^b}$$_{[h_1(1170)^0]}$                 &                                    \\
                                                       &$^{0.066\pm0.066^a}_{0.0050\pm0.0050^b}$$_{[h_1(1415)^0]}$         &                                              &$^{0.042\pm0.042^a}_{0.0029\pm0.0029^b}$$_{[h_1(1415)^0]}$               &                                    \\\hline
$B^+\to \phi\eta\tau^+\nu_\tau$                        &$^{0.010\pm0.010^a}_{0.0088\pm0.0088^b}$$_{[h_1(1170)^0]}$         &                                              &$^{0.0019\pm0.0019^a}_{0.0019\pm0.0019^b}$$_{[h_1(1170)^0]}$               &                                    \\
                                                       &$^{0.010\pm0.010^a}_{0.0014\pm0.0014^b}$$_{[h_1(1415)^0]}$         &                                              &$^{0.0019\pm0.0019^a}_{0.00025\pm0.00025^b}$$_{[h_1(1415)^0]}$             &                                    \\\hline
$B^+\to \omega\pi^0\tau^+\nu_\tau$                     &$8.38\pm3.00_{[b_1(1235)^0]}$                                      &                                              &$7.18\pm2.31_{[b_1(1235)^0]}$                 &                                    \\\hline
$B^+\to \rho^0\eta\tau^+\nu_\tau$                      &$0.67\pm0.26_{[b_1(1235)^0]}$                                      &                                              &$0.18\pm0.07_{[b_1(1235)^0]}$                 &                                    \\\hline
$B^+\to K^{*+}K^-\tau^+\nu_\tau$                       &$0.13\pm0.05_{[a_1(1260)^0]}$                                      &$0.000053\pm0.000035_{[a_2(1320)^0]}$           &$0.045\pm0.019_{[a_1(1260)^0]}$                 &$0.000022\pm0.000014_{[a_2(1320)^0]}$        \\
or                                                     &$0.057\pm0.022_{[b_1(1235)^0]}$                                    &$0.000080\pm0.000055_{[f_2(1270)^0]}$           &$0.025\pm0.009_{[b_1(1235)^0]}$                 &$0.000024\pm0.000016_{[f_2(1270)^0]}$        \\
$B^+\to K^{*-}K^+\tau^+\nu_\tau$                       &$^{0.019\pm0.018^a}_{0.032\pm0.018^b}$$_{[f_1(1285)^0]}$           &$0.000079\pm0.000060_{[f_2'(1525)^0]}$          &$0_{[f_1(1285)^0]}$               &$0.000059\pm0.000044_{[f_2'(1525)^0]}$       \\
                                                       &$^{0.89\pm89^a}_{0.19\pm0.19^b}$$_{[f_1(1420)^0]}$                 &                                              &$^{0.49\pm0.49^a}_{0.10\pm0.10^b}$$_{[f_1(1420)^0]}$                 &                                           \\
                                                       &$^{0.13\pm0.09^a}_{0.14\pm0.10^b}$$_{[h_1(1170)^0]}$               &                                              &$^{0.031\pm0.022^a}_{0.037\pm0.025^b}$$_{[h_1(1170)^0]}$                 &                                           \\
                                                       &$^{0.056\pm0.056^a}_{0.016\pm0.016^b}$$_{[h_1(1415)^0]}$           &                                              &$^{0.040\pm0.040^a}_{0.0099\pm0.0099^b}$$_{[h_1(1415)^0]}$               &                                           \\\hline
$B^+\to K^{*0}\overline{K}^0\tau^+\nu_\tau$            &$0.13\pm0.05_{[a_1(1260)^0]}$                                      &$0.000042\pm0.000028_{[a_2(1320)^0]}$           &$0.043\pm0.017_{[a_1(1260)^0]}$                 &$0.000018\pm0.000011_{[a_2(1320)^0]}$        \\
or                                                     &$0.050\pm0.020_{[b_1(1235)^0]}$                                    &$0.000071\pm0.000049_{[f_2(1270)^0]}$           &$0.022\pm0.008_{[b_1(1235)^0]}$                 &$0.000022\pm0.000014_{[f_2(1270)^0]}$        \\
$B^+\to \overline{K}^{*0}K^0\tau^+\nu_\tau$            &$^{0.0047\pm0.0048^a}_{0.0075\pm0.0044^b}$$_{[f_1(1285)^0]}$       &$0.000071\pm0.000053_{[f_2'(1525)^0]}$          &$0_{[f_1(1285)^0]}$               &$0.000053\pm0.000039_{[f_2'(1525)^0]}$       \\
                                                       &$^{0.80\pm0.80^a}_{0.17\pm0.17^b}$$_{[f_1(1420)^0]}$               &                                              &$^{0.43\pm0.43^a}_{0.09\pm0.09^b}$$_{[f_1(1420)^0]}$                 &                                           \\
                                                       &$^{0.12\pm0.09^a}_{0.13\pm0.10^b}$$_{[h_1(1170)^0]}$               &                                              &$^{0.029\pm0.021^a}_{0.035\pm0.023^b}$$_{[h_1(1170)^0]}$               &                                           \\
                                                       &$^{0.047\pm0.047^a}_{0.013\pm0.013^b}$$_{[h_1(1415)^0]}$           &                                              &$^{0.035\pm0.035^a}_{0.0088\pm0.0088^b}$$_{[h_1(1415)^0]}$               &                                           \\\hline
$B^0\to \rho^0\pi^-\tau^+\nu_\tau$                     &$5.17\pm2.21_{[a_1(1260)^-]}$                                      &$0.062\pm0.039_{[a_2(1320)^-]}$                 &$2.64\pm1.06_{[a_1(1260)^-]}$                 &$0.052\pm0.033_{[a_2(1320)^-]}$                    \\\hline
$B^0\to \rho^-\pi^0\tau^+\nu_\tau$                     &$5.20\pm2.22_{[a_1(1260)^-]}$                                      &$0.062\pm0.040_{[a_2(1320)^-]}$                 &$2.65\pm1.04_{[a_1(1260)^-]}$                 &$0.053\pm0.033_{[a_2(1320)^-]}$                    \\\hline
$B^0\to \omega\pi^-\tau^+\nu_\tau$                     &$15.48\pm5.53_{[b_1(1235)^-]}$                                     &                                              &$2.88\pm0.92_{[b_1(1235)^-]}$                &                                                 \\\hline
$B^0\to \rho^-\eta\tau^+\nu_\tau$                      &$1.25\pm0.50_{[b_1(1235)^-]}$                                      &                                              &$0.23\pm0.09_{[b_1(1235)^-]}$                 &                                                 \\\hline
$B^0\to K^{*0}K^-\tau^+\nu_\tau$                       &$0.49\pm0.20_{[a_1(1260)^-]}$                                      &$0.00017\pm0.00011_{[a_2(1320)^-]}$           &$0.16\pm0.07_{[a_1(1260)^-]}$                 &$0.000073\pm0.000047_{[a_2(1320)^-]}$              \\
                                                       &$0.20\pm0.08_{[b_1(1235)^-]}$                                      &                                              &$0.088\pm0.036_{[b_1(1235)^-]}$                 &                                                 \\\hline
$B^0\to K^{*-}K^0\tau^+\nu_\tau$                       &$0.49\pm0.20_{[a_1(1260)^-]}$                                      &$0.00018\pm0.00012_{[a_2(1320)^-]}$           &$0.16\pm0.07_{[a_1(1260)^-]}$                 &$0.000073\pm0.000047_{[a_2(1320)^-]}$              \\
                                                       &$0.20\pm0.08_{[b_1(1235)^-]}$                                      &                                              &$0.087\pm0.036_{[b_1(1235)^-]}$                 &                                                 \\\hline
$B_s^0\to K^{*-}\pi^0\tau^+\nu_\tau$                   &$^{2.37\pm1.45^\natural}_{2.53\pm1.57^\sharp}$$_{[K_1(1270)^-]}$               &$0.013\pm0.008_{[K_2^*(1430)^-]}$             &$^{1.93\pm1.07^\natural}_{2.01\pm1.13^\sharp}$$_{[K_1(1270)^-]}$                 &$0.011\pm0.007_{[K_2^*(1430)^-]}$                \\
                                                       &$^{0.86\pm0.60^\natural}_{0.36\pm0.28^\sharp}$$_{[K_1(1400)^-]}$               &                                              &$^{0.57\pm0.39^\natural}_{0.31\pm0.23^\sharp}$$_{[K_1(1400)^-]}$                 &                                                 \\\hline
$B_s^0\to \overline{K}^{*0}\pi^-\tau^+\nu_\tau$        &$^{4.66\pm2.85^\natural}_{4.95\pm3.09^\sharp}$$_{[K_1(1270)^-]}$               &$0.024\pm0.016_{[K_2^*(1430)^-]}$             &$^{3.78\pm2.09^\natural}_{3.92\pm2.19^\sharp}$$_{[K_1(1270)^-]}$                 &$0.021\pm0.014_{[K_2^*(1430)^-]}$                \\
                                                       &$^{1.71\pm1.19^\natural}_{0.71\pm0.55^\sharp}$$_{[K_1(1400)^-]}$               &                                              &$^{1.12\pm0.77^\natural}_{0.60\pm0.45^\sharp}$$_{[K_1(1400)^-]}$                 &                                                 \\\hline
$B_s^0\to \rho^0K^-\tau^+\nu_\tau$                     &$^{5.34\pm1.63^\natural}_{5.62\pm1.62^\sharp}$$_{[K_1(1270)^-]}$               &$0.0044\pm0.0029_{[K_2^*(1430)^-]}$             &$^{2.32\pm0.64^\natural}_{2.45\pm0.61^\sharp}$$_{[K_1(1270)^-]}$                 &$0.0046\pm0.0030_{[K_2^*(1430)^-]}$                \\
                                                       &$^{0.031\pm0.031^\natural}_{0.0085\pm0.0085^\sharp}$$_{[K_1(1400)^-]}$         &                                              &$^{0.016\pm0.016^\natural}_{0.0064\pm0.0064^\sharp}$$_{[K_1(1400)^-]}$                 &                                                 \\\hline
$B_s^0\to \rho^-\overline{K}^0\tau^+\nu_\tau$          &$^{10.33\pm3.11^\natural}_{10.92\pm3.16^\sharp}$$_{[K_1(1270)^-]}$             &$0.0086\pm0.0057_{[K_2^*(1430)^-]}$             &$^{4.29\pm1.17^\natural}_{4.51\pm1.12^\sharp}$$_{[K_1(1270)^-]}$                 &$0.0087\pm0.0058_{[K_2^*(1430)^-]}$                \\
                                                       &$^{0.061\pm0.061^\natural}_{0.017\pm0.017^\sharp}$$_{[K_1(1400)^-]}$           &                                              &$^{0.031\pm0.031^\natural}_{0.013\pm0.013^\sharp}$$_{[K_1(1400)^-]}$                 &                                                 \\\hline
$B_s^0\to \omega K^-\tau^+\nu_\tau$                    &$^{2.97\pm0.89^\natural}_{3.11\pm0.89^\sharp}$$_{[K_1(1270)^-]}$               &$0.0049\pm0.0036_{[K_2^*(1430)^-]}$             &$^{1.99\pm0.54^\natural}_{2.10\pm0.53^\sharp}$$_{[K_1(1270)^-]}$                 &$0.0051\pm0.0038_{[K_2^*(1430)^-]}$                \\
                                                       &$^{0.030\pm0.030^\natural}_{0.0082\pm0.0082^\sharp}$$_{[K_1(1400)^-]}$         &                                              &$^{0.015\pm0.015^\natural}_{0.0062\pm0.0062^\sharp}$$_{[K_1(1400)^-]}$                 &                                                 \\\hline
$B_s^0\to K^{*-}\eta\tau^+\nu_\tau$                    &$^{0.17\pm0.09^\natural}_{0.17\pm0.09^\sharp}$$_{[K_1(1270)^-]}$               &$0.00089\pm0.00060_{[K_2^*(1430)^-]}$           &$^{0.056\pm0.033^\natural}_{0.057\pm0.033^\sharp}$$_{[K_1(1270)^-]}$                 &$0.00041\pm0.00027_{[K_2^*(1430)^-]}$              \\
                                                       &$^{0.079\pm0.052^\natural}_{0.043\pm0.036^\sharp}$$_{[K_1(1400)^-]}$           &                                              &$^{0.034\pm0.022^\natural}_{0.028\pm0.024^\sharp}$$_{[K_1(1400)^-]}$                 &                                                 \\\hline
$B_s^0\to \phi K^-\tau^+\nu_\tau$                      & $^{0.11\pm0.08^\natural}_{0.048\pm0.037^\sharp}$$_{[K_1(1400)^-]}$            &$0.000059\pm0.000041_{[K_2^*(1430)^-]}$         &$^{0.044\pm0.03^\natural}_{0.027\pm0.020^\sharp}$$_{[K_1(1400)^-]}$                                               &$0.000032\pm0.000021_{[K_2^*(1430)^-]}$             \\\hline
\end{tabular}}\label{Tab:B2PVlv2}
\end{table}

From the second and third columns of  Tabs. \ref{Tab:B2PVlv1} and \ref{Tab:B2PVlv2}, one can see that some branching ratios with the axial-vector meson resonances,
such as $\mathcal{B}(B^+\to \rho^\mp\pi^\pm\ell^+\nu_{\ell})_{[a_1(1260)^0,h_1(1170)^0]}$, $\mathcal{B}(B^+\to \rho^0\pi^0\ell^+\nu_{\ell})_{[h_1(1170)^0]}$,  $\mathcal{B}(B^+\to \omega\pi^0\ell^+\nu_{\ell})_{[b_1(1235)^0]}$,  $\mathcal{B}(B^+\to \rho^0\eta\ell'^+\nu_{\ell'})_{[b_1(1235)^0]}$, $\mathcal{B}(B^0\to \rho^0\pi^-\ell^+\nu_{\ell})_{[a_1(1260)^-]}$, $\mathcal{B}(B^0\to \rho^-\pi^0\ell^+\nu_{\ell})_{[a_1(1260)^-]}$, $\mathcal{B}(B^0\to \omega\pi^-\ell^+\nu_{\ell})_{[b_1(1235)^-]}$, $\mathcal{B}(B^0\to \rho^-\eta\ell^+\nu_{\ell})_{[b_1(1235)^-]}$, $\mathcal{B}(B^0_s\to K^{*-}\pi^0\ell^+\nu_{\ell})_{[K_1(1270)^-]}$, $\mathcal{B}(B^0_s\to \overline{K}^{*0}\pi^-\ell^+\nu_{\ell})_{[K_1(1270)^-,K_1(1400)^-]}$, $\mathcal{B}(B^0_s\to \rho^0K^-\ell^+\nu_{\ell})_{[K_1(1270)^-]}$, $\mathcal{B}(B^0_s\to \rho^-\overline{K}^0\ell^+\nu_{\ell})_{[K_1(1270)^-]}$ and  $\mathcal{B}(B^0_s\to \omega K^-\ell^+\nu_{\ell})_{[K_1(1270)^-]}$,
 are  on the order of $\mathcal{O}(10^{-4}-10^{-3})$, and other branching ratios with the axial-vector meson resonances and all branching ratios with the tensor meson resonances are small.

\subsection{ $\mathcal{B}(B\to PV\ell^+\nu_{\ell})$ with the  width effects }
 Some decay widths of the resonance  states, such as ones of $h_1(1170)$, $h_1(1415)$, $b_1(1235)$,  $a_1(1260)$,  $K_1(1270)$, $K_1(1400)$,  $f_2(1270)$, $f'_2(1270)$, $a_2(1320)$, and $K_2(1400)$,
 are not very narrow.  We will consider the width effects of the resonance  states.
 After considering the width effects of the resonance  states,
the decay branching ratios of $B\to P V\ell^+\nu_\ell$ are \cite{Cheng:1993ah,Tsai:2021ota}
{\small
\begin{eqnarray}
\mathcal{B}(B\to PV \ell^+\nu_\ell)_{R}=\frac{1}{\pi}\int_{(m_{R}-n\Gamma_{R})^2}^{(m_{R}+n\Gamma_{R})^2} dt_R \int_{m^2_{\ell}}^{(m_B-\sqrt{t_R})^2} dq^2\frac{\sqrt{t_R}d\mathcal{B}(B\to R\ell^+\nu_\ell,t_R)/dq^2~\mathcal{B}(R\to PV,t_R)\Gamma_{R}}{(t_R-m_{R}^2)^2+m^2_{R}\Gamma^2_{R}},\label{Eq:RDA4}
\end{eqnarray}}
where $d\mathcal{B}(B\to R\ell^+\nu_\ell,t_R)/dq^2$ and $\mathcal{B}(R\to PV,t_R)$ are obtained from  $d\mathcal{B}(B\to R\ell^+\nu_\ell)/dq^2$ and $\mathcal{B}(R\to PV)$ by replacing $m_R \to \sqrt{t_R}$.
The form factors of the $B\to R\ell^+\nu_\ell$ decays and the coefficients of the R-V-P vertexes follow the SU(3) flavor symmetry relations.  And then the predicted branching ratios are obtained, which are listed in the last two columns of Tabs.  \ref{Tab:B2PVlv1} and \ref{Tab:B2PVlv2} and  named as ones in the $S_2$ case.
Comparing with the results with the narrow width  approximation, one can see that some results  considered the resonant width effects are obviously  smaller.
For $\mathcal{B}(B^+\to K^{*\pm}K^{\mp} \ell^+\nu_\ell)_{[f_1(1285)^0]}$,  $\mathcal{B}(B^+\to K^{*0}\overline{K}^{0} \ell^+\nu_\ell)_{[f_1(1285)^0]}$ and $\mathcal{B}(B^+\to \overline{K}^{*0}K^{0} \ell^+\nu_\ell)_{[f_1(1285)^0]}$,
they are not allowed in the $S_2$ case, since the decay width of $f_1(1285)^0$ is small. Note  that  for  $f_1(1285)^0\to K^{*+}K^{-},K^{*-}K^{+},K^{*0}\overline{K}^{0},\overline{K}^{*0}K^{0}$ decays, their  phase spaces are very narrow, and their branching ratios have been obtained by considering the decay widths of both $f_1(1285)^0$ and $K^{*}$ mesons in Ref. \cite{Wang:2022fbk}. However, we only consider the decay width of $f_1(1285)^0$ meson in the $S_2$ case, and this is the reason that the predictions are different between the $S_1$ and $S_2$ cases.

\subsection{ $\mathcal{B}(B\to PV\ell^+\nu_{\ell})$ with the excited vector resonance states }
In addition, the excited vector meson contributions to the branching ratios of  four-body semi-leptonic decays will be calculated by  the narrow width approximation in this subsection.
Since $\mathcal{B}(B\to V_{E,D}\ell^+\nu_{\ell})$ have been listed in Tab. \ref{Tab:BrB2VEDlv}, now we just need to calculate $\mathcal{B}(V_{E,D}\to PV)$ to get $\mathcal{B}(B\to PV\ell^+\nu_{\ell})$ with the excited vector resonance states.

The branching ratios of the $V_{E,D}\to PV$ strong decays can be written as  \cite{Liang:2024lon}
\begin{eqnarray}
\mathcal{B}(V'\to PV)=\frac{\tau_{V'}p^3_c}{12\pi m^2_{V'}}g^2_{V'\to PV},
\end{eqnarray}
where $V'=V_E$ or $V_D$,  $p_c\equiv\frac{\sqrt{\lambda(m_{V'}^2,m_{P}^2,m_{V}^2)}}{2m_{V'}}$  and  $g_{V'\to PV}$ are the strong coupling constants.    $g_{V'\to PV}$  can be  parametrized  by  the SU(3) flavor symmetry
\begin{eqnarray}
g_{V'\to PV}=g_{V'}V'^i_jP^k_iV^j_k, \label{Eq:gV2PP0}
\end{eqnarray}
where $g_{V'}$ are the corresponding nonperturbative parameters. The strong coupling constants
of these decays are listed in  Tab. \ref{Tab:StronggVED2PV}.

For the $V_D\to PV$ strong decays, only $K^*(1680)\to \rho K, K^*\pi$ decays have been measured \cite{PDG2024}], as follows:
\begin{eqnarray}
\mathcal{B}(K^*(1680)\to \rho K)^{Exp}=(31.4^{+5.0}_{-2.1})\%,~~~~~~~~\mathcal{B}(K^*(1680)\to K^*\pi)^{Exp}=(29.9^{+2.2}_{-5.0})\%,\label{Eq:VE2PVEXP}
\end{eqnarray}
\begin{table}[h]
\renewcommand\arraystretch{1.0}
\tabcolsep 0.06in
\centering
\caption{Strong coupling constants of the $V'\to PV$ decays by the SU(3) flavor symmetry.}\vspace{0.08cm}
{\footnotesize
\begin{tabular}{lc|lc}  \hline\hline
~~~~~~~~~~~~Decay modes &  Strong couplings $g_{V'\to PV}$                                             &~~~~~~~~~~~~Decay modes &  Strong couplings $g_{V'\to PV}$\\\hline
$K^{*}(1410)^+/K^{*}(1680)^+\to \rho^0K^+$                &  $\frac{1}{\sqrt{2}}g_{V'}$                                                         &   $K^{*}(1410)^0/K^{*}(1680)^0\to \rho^-K^+$                &  $g_{V'}$                                                                       \\
$K^{*}(1410)^+/K^{*}(1680)^+\to \omega K^+$               &  $\frac{1}{\sqrt{2}}g_{V'}$                                                         &   $K^{*}(1410)^0/K^{*}(1680)^0\to \rho^0K^0$                &  $\frac{1}{\sqrt{2}}g_{V'}$                                                     \\
$K^{*}(1410)^+/K^{*}(1680)^+\to \rho^+ \overline{K}^0$    &  $g_{V'}$                                                                           &   $K^{*}(1410)^0/K^{*}(1680)^0\to \omega K^0$               &  $\frac{1}{\sqrt{2}}g_{V'}$                                                     \\
$K^{*}(1410)^+/K^{*}(1680)^+\to \phi K^+$                 &  $g_{V'}$                                                                           &   $K^{*}(1410)^0/K^{*}(1680)^0\to \phi K^0$                 &  $g_{V'}$                                                                        \\
$K^{*}(1410)^+/K^{*}(1680)^+\to K^{*+} \pi^0$             &  $\frac{1}{\sqrt{2}}g_{V'}$                                                         &   $K^{*}(1410)^0/K^{*}(1680)^0\to K^{*+} \pi^-$             &  $g_{V'}$                                                                        \\
$K^{*}(1410)^+/K^{*}(1680)^+\to \overline{K}^{*0} \pi^+$  &  $g_{V'}$                                                                           &   $K^{*}(1410)^0/K^{*}(1680)^0\to K^{*0} \pi^0$             &  $-\frac{1}{\sqrt{2}}g_{V'}$                                                                  \\
$K^{*}(1410)^+/K^{*}(1680)^+\to K^{*+}\eta$    &  $\big(\frac{cos\theta_P}{\sqrt{6}}-\frac{sin\theta_P}{\sqrt{3}}\big)g_{V'}$                   &   $K^{*}(1410)^0/K^{*}(1680)^0\to K^{*0}\eta$               &  $\big(\frac{cos\theta_P}{\sqrt{6}}-\frac{sin\theta_P}{\sqrt{3}}\big)g_{V'}$                \\
$K^{*}(1410)^+/K^{*}(1680)^+\to K^{*+}\eta'$   &  $\big(\frac{sin\theta_P}{\sqrt{6}}+\frac{cos\theta_P}{\sqrt{3}}\big)g_{V'}$                   &   $K^{*}(1410)^0/K^{*}(1680)^0\to K^{*0}\eta'$              &  $\big(\frac{sin\theta_P}{\sqrt{6}}+\frac{cos\theta_P}{\sqrt{3}}\big)g_{V'}$                \\\hline
$\rho(1450)^0/\rho(1700)^0\to \rho^0\eta$                 &  $\big(\frac{2cos\theta_P}{\sqrt{6}}-\frac{2sin\theta_P}{\sqrt{3}}\big)g_{V'}$       &  $\rho(1450)^+/\rho(1700)^+\to \rho^+\eta$                 &  $\big(\frac{2cos\theta_P}{\sqrt{6}}-\frac{2sin\theta_P}{\sqrt{3}}\big)g_{V'}$      \\
$\rho(1450)^0/\rho(1700)^0\to \rho^0\eta'$                &  $\big(\frac{2sin\theta_P}{\sqrt{6}}+\frac{2cos\theta_P}{\sqrt{3}}\big)g_{V'}$       &  $\rho(1450)^+/\rho(1700)^+\to \rho^+\eta'$                &  $\big(\frac{2sin\theta_P}{\sqrt{6}}+\frac{2cos\theta_P}{\sqrt{3}}\big)g_{V'}$      \\
$\rho(1450)^0/\rho(1700)^0\to \omega\pi^0$                &  $\sqrt{2}g_{V'}$                                                                    &  $\rho(1450)^+/\rho(1700)^+\to \omega\pi^+$                &  $\sqrt{2}g_{V'}$                                                                   \\
$\rho(1450)^0/\rho(1700)^0\to K^{*+}K^-$                  &  $\frac{1}{\sqrt{2}}g_{V'}$                                                          &  $\rho(1450)^+/\rho(1700)^+\to K^{*+}\overline{K}^0$       &  $g_{V'}$                                               \\
$\rho(1450)^0/\rho(1700)^0\to K^{*-}K^+$                  &  $\frac{1}{\sqrt{2}}g_{V'}$                                                          &  $\rho(1450)^+/\rho(1700)^+\to \overline{K}^{*0}K^+$       &  $g_{V'}$                                               \\
$\rho(1450)^0/\rho(1700)^0\to \overline{K}^{*0}K^0$       &  $-\frac{1}{\sqrt{2}}g_{V'}$                                                         &                                                                                                                                                     \\
$\rho(1450)^0/\rho(1700)^0\to K^{*0}\overline{K}^0$       &  $-\frac{1}{\sqrt{2}}g_{V'}$                                                         &                                                                                                                                                   \\ \hline
$\omega(1420)/\omega(1650)\to \rho^0\eta$                 &  $\big(\frac{2cos\theta_P}{\sqrt{6}}-\frac{2sin\theta_P}{\sqrt{3}}\big)g_{V'}$       &  $\omega(1420)/\omega(1650)\to K^{*+}K^-$                   &  $\frac{1}{\sqrt{2}}g_{V'}$                                                       \\
$\omega(1420)/\omega(1650)\to \rho^0\eta'$                &  $\big(\frac{2sin\theta_P}{\sqrt{6}}+\frac{2cos\theta_P}{\sqrt{3}}\big)g_{V'}$       &  $\omega(1420)/\omega(1650)\to K^{*-}K^+$                   &  $\frac{1}{\sqrt{2}}g_{V'}$                                                        \\
$\omega(1420)/\omega(1650)\to \rho^0\pi^0$                &  $\sqrt{2}g_{V'}$                                                                    &  $\omega(1420)/\omega(1650)\to \overline{K}^{*0}K^0$        &  $\frac{1}{\sqrt{2}}g_{V'}$                                                         \\
$\omega(1420)/\omega(1650)\to \rho^+\pi^-$                &  $\sqrt{2}g_{V'}$                                                                    &  $\omega(1420)/\omega(1650)\to K^{*0}\overline{K}^0$        &  $\frac{1}{\sqrt{2}}g_{V'}$                                                          \\
$\omega(1420)/\omega(1650)\to \rho^-\pi^+$                &  $\sqrt{2}g_{V'}$                                                                                                                                                                                                                           \\\hline
\end{tabular}}\label{Tab:StronggVED2PV}
\renewcommand\arraystretch{0.8}
\tabcolsep 0.05in
\centering
\caption{Branching ratios and decay widths of the $V_D\to PV$ strong decays by the SU(3) flavor symmetry.}\vspace{0.08cm}
{\footnotesize
\begin{tabular}{lcc|lcc}  \hline\hline
~~$ V_D\to PV$ decays                        &  Branching ratios [\%]   & Decay width [MeV]                &~~$ V_D\to PV$ decays &  Branching ratios [\%]   & Decay width [MeV]\\\hline
$K^{*}(1680)^+\to \rho^0K^+$                &  $5.90\pm1.43$         & $21.61\pm2.63$                    &   $K^{*}(1680)^0\to \rho^-K^+$                &$11.80\pm2.86$       &  $43.25\pm5.27$       \\
$K^{*}(1680)^+\to \omega K^+$               &  $5.76\pm1.40$         & $21.11\pm2.59$                    &   $K^{*}(1680)^0\to \rho^0K^0$                &$5.84\pm1.42$        & $21.39\pm2.61$        \\
$K^{*}(1680)^+\to \rho^+ \overline{K}^0$    &  $11.68\pm2.83$        &  $42.81\pm5.23$                   &   $K^{*}(1680)^0\to \omega K^0$               &$5.70\pm1.39$        & $20.90\pm2.57$         \\
$K^{*}(1680)^+\to \phi K^+$                 &  $3.72\pm1.13$         &  $13.39\pm2.45$                   &   $K^{*}(1680)^0\to \phi K^0$                 &$3.63\pm1.11$        &  $13.07\pm2.42$        \\
$K^{*}(1680)^+\to K^{*+} \pi^0$             &  $7.55\pm1.82$         & $27.44\pm3.03$                    &   $K^{*}(1680)^0\to K^{*+} \pi^-$             &$15.06\pm3.63$       & $54.70\pm6.04$         \\
$K^{*}(1680)^+\to \overline{K}^{*0} \pi^+$  &  $14.90\pm3.59$        & $54.15\pm6.01$                    &   $K^{*}(1680)^0\to K^{*0} \pi^0$             &$7.48\pm1.80$        &  $27.16\pm3.01$        \\
$K^{*}(1680)^+\to K^{*+}\eta$               &  $1.87\pm0.72$         & $6.69\pm1.83$                     &   $K^{*}(1680)^0\to K^{*0}\eta$               &$1.84\pm0.71$        &  $6.54\pm1.79$          \\
$K^{*}(1680)^+\to K^{*+}\eta'$              &  $\cdots$              &  $\cdots$                         &   $K^{*}(1680)^0\to K^{*0}\eta'$              &$\cdots$             &  $\cdots$                        \\
$K^{*}(1680)\to K^{*}\pi$                   &  $22.50\pm5.41$        & $81.74\pm9.03$                    &   $K^{*}(1680)\to K^{*}\eta$                  &$1.85\pm0.71$        &$6.62\pm1.81$             \\
$K^{*}(1680)\to \rho K$                     &  $17.61\pm4.27$        & $64.53\pm7.87$                    &   $K^{*}(1680)\to \omega K$                   &$5.73\pm1.40$        & $21.00\pm2.58$           \\
$K^{*}(1680)\to \phi K$                     &  $3.67\pm1.12$         & $13.23\pm2.43$                                                                                                                        \\ \hline
$\rho(1700)^0\to \rho^0\eta$                 &$14.99\pm5.41$         & $44.08\pm10.83$                    &  $\rho(1700)^+\to \rho^+\eta$                 &$14.99\pm5.41$       &  $44.11\pm10.83$           \\
$\rho(1700)^0\to \rho^0\eta'$                &$0.020\pm0.020$        & $0.048\pm0.048$                    &  $\rho(1700)^+\to \rho^+\eta'$                &$0.020\pm0.020$      &   $0.050\pm0.050$          \\
$\rho(1700)^0\to \omega\pi^0$                &$48.50\pm12.03$        & $141.91\pm15.14$                   &  $\rho(1700)^+\to \omega\pi^+$                &$48.37\pm12.00$      & $141.56\pm15.12$           \\
$\rho(1700)^0\to K^{*+}K^-$                  &$4.77\pm1.31$          & $14.09\pm2.05$                     &  $\rho(1700)^+\to K^{*+}\overline{K}^0$       &$9.42\pm2.59$        &   $27.78\pm4.06$           \\
$\rho(1700)^0\to K^{*-}K^+$                  &$4.77\pm1.31$          & $14.08\pm2.05$                     &  $\rho(1700)^+\to \overline{K}^{*0}K^+$       &$9.39\pm2.59$        &  $27.69\pm4.06$             \\
$\rho(1700)^0\to \overline{K}^{*0}K^0$       &$4.63\pm1.28$          & $13.65\pm2.01$                     &  $\rho(1700)\to \rho\eta$                     &$14.99\pm5.41$       & $44.09\pm10.83$              \\
$\rho(1700)^0\to K^{*0}\overline{K}^0$       &$4.63\pm1.28$          & $13.65\pm2.01$                     &  $\rho(1700)\to \rho\eta'$                    &$0.020\pm0.020$      & $0.049\pm0.049$              \\
$\rho(1700)\to \omega\pi$                    &$48.43\pm12.02$        & $141.74\pm15.13$                   & $\rho(1700)\to K^*K$                          &$18.81\pm5.19$       &  $55.48\pm8.11$              \\ \hline
$\omega(1650)\to \omega\eta$                 &$10.13\pm2.51$         & $34.84\pm9.12$                     &  $\omega(1650)\to K^{*+}K^-$                  & $3.11\pm0.50$        & $10.80\pm1.73$            \\
$\omega(1650)\to \omega\eta'$                & $\cdots$              & $\cdots$                           &  $\omega(1650)\to K^{*-}K^+$                  & $3.11\pm0.50$        & $10.80\pm1.73$             \\
$\omega(1650)\to \rho^0\pi^0$                &$36.43\pm2.98$         & $126.71\pm10.30$                   &  $\omega(1650)\to \overline{K}^{*0}K^0$       & $3.00\pm0.49$        & $10.41\pm1.72$              \\
$\omega(1650)\to \rho^+\pi^-$                &$36.32\pm2.97$         & $126.33\pm10.27$                   &  $\omega(1650)\to K^{*0}\overline{K}^0$       & $3.00\pm0.49$        & $10.41\pm1.72$               \\
$\omega(1650)\to \rho^-\pi^+$                &$36.32\pm2.97$         & $126.33\pm10.27$                   &  $\omega(1650)\to K^*K$                       & $12.21\pm1.97$       & $42.42\pm6.89$               \\
$\omega(1650)\to \rho\pi$                    &$108.97\pm8.90$        & $378.99\pm30.82$                                                                                                                          \\\hline
\end{tabular}}\label{Tab:VD2PVBr}
\end{table}
and they can be used to constrain the nonperturbative parameter $g_{V_D}$. We obtain $g_{V_D}=6.48\pm1.59$ from $\mathcal{B}(K^*(1680)\to K^*\pi)^{Exp}$ and $g_{V_D}=5.33\pm1.33$ from $\mathcal{B}(K^*(1680)\to K^*\pi)^{Exp}$.
The communication between them, $g_{V_D}\in[4.88,6.66]$, will be used to obtained the branching ratios and decay widths of the $V_D\to PV$ strong decays.  After considering the bounds of $\mathcal{B}(K^*(1680)\to PV)\leq63.2\%$,
$\mathcal{B}(\rho(1700)\to PV)\leq100\%$, and $\mathcal{B}(\omega(1650)\to PV)\leq130\%$,  the predictions are obtained and listed in Tab. \ref{Tab:VD2PVBr}.
Please note that $\mathcal{B}(\omega(1650)\to PV)$, which is defined as $\mathcal{B}(\omega(1650)\to \rho\pi)+\mathcal{B}(\omega(1650)\to K^*K)+\mathcal{B}(\omega(1650)\to \omega\eta)$,  is always larger than 1, so we have to extend that $\mathcal{B}(\omega(1650)\to PV)\leq130\%$,  which means that   all branching ratio  predictions of every $\omega(1650)\to PV$ modes should be smaller than our predictions.  Moreover,  we set $\mathcal{B}(K^*(1680)\to PV)\leq63.2\%$, since $\mathcal{B}(K^*(1680)\to K\pi)=(38.7\pm2.5)\%$ and $\mathcal{B}(K^*(1680)\to K\eta)=(1.4^{+1.0}_{-0.8})\%$  are deducted \cite{PDG2024}.

From Tab. \ref{Tab:VD2PVBr}, one can see that the errors of some predictions are  large, and  the errors mainly come from the  nonperturbative parameter $g_{V_D}$  and  the decay widths such as $ \Gamma_{K^*(1680)}=320\pm110$ MeV,
$\Gamma_{\rho(1700)}=250\pm100$ MeV and $\Gamma_{\omega(1650)}=315\pm35$ MeV \cite{PDG2024}. Our prediction of $\mathcal{B}(K^{*}(1680)\to K^{*}\pi)$ is consistent with its experimental data within $1\sigma$ error,
 nevertheless,   our prediction of $\mathcal{B}(K^{*}(1680)\to \rho K)$ is smaller than its experimental data,
since we only use the experimental data to obtain  $g_{V_D}\in[4.88,6.66]$, but do not use the data given in Eq. (\ref{Eq:VE2PVEXP}) to further constrain the predictions.
The reason is that  within the $1\sigma$ error of the theoretical inputs and the experimental data, it is difficult  to satisfy $\mathcal{B}(K^{*}(1680)\to \rho K)^{Exp}$  and $\mathcal{B}(K^{*}(1680)\to K^{*}\pi)^{Exp}$ at the same time.
If we extend the experimental errors to 2$\sigma$,
the experimental data of $\mathcal{B}(K^*(1680)\to\rho K)$ and $\mathcal{B}(K^*(1680)\to  K^* \pi)$  can be satisfied at the same time, and the errors of the $\mathcal{B}(K^*(1680)\to PV)$ predictions are very small, for example:
\begin{eqnarray}
\mathcal{B}(K^{*}(1680)\to K^{*}\pi)&=&(34.12\pm0.18)\%,  ~~~~~~~~~\mathcal{B}(K^{*}(1680)\to \rho K)=(27.34\pm0.14)\%, \nonumber\\
\mathcal{B}(K^{*}(1680)\to \omega K)&=&(8.90\pm0.05)\%,   ~~~~~~~~~~\mathcal{B}(K^{*}(1680)\to K^{*}\eta)=(2.85\pm0.44)\%,\nonumber\\
\mathcal{B}(K^{*}(1680)\to \phi K)&=&(5.89\pm0.18)\%.
\end{eqnarray}
Nevertheless, other predictions including $\Gamma(K^(1680)\to PV)$ still contain large errors.
The branching ratio predictions in Tab. \ref{Tab:VD2PVBr} will be used to obtain  $\mathcal{B}(B\to PV\ell^+\nu_{\ell})_{[V_D]}$ later.

Strong decays of the excited vector mesons  have been studied by using  effective relativistic quantum field theoretical model based on flavor symmetry \cite{Piotrowska:2017rgt},
 in which the charges of the final state particle are not distinguished. For convenient comparison with them, our decay width predictions are also listed in Tab. \ref{Tab:VD2PVBr}.
 Our prediction of $\Gamma(K^{*}(1680)\to K^{*}\eta)$ is 1 order of magnitude larger than the one in Ref. \cite{Piotrowska:2017rgt}.  Our predictions of  other decay widths of the $V_D\to PV$ decays are on  the same order of magnitude as the  ones in Ref. \cite{Piotrowska:2017rgt} with smaller errors.

As for the $V_E\to PV$ decays, no  modes have been measured at present, and there are only two limits \cite{PDG2024}, as follows:
\begin{eqnarray}
\mathcal{B}(K^{*}(1410)\to K^{*}\pi)>40\%,  ~~~~~\mathcal{B}(K^{*}(1410)\to \rho K)<7\%, ~~~~~\mbox{at the} ~~95\%~~\mbox{CL}
\end{eqnarray}
and they cannot be satisfied at same time within the SU(3) flavor symmetry.
Therefore, $g_{V_E}=\frac{18.4\pm3.8}{16.5\pm3.5}g_{V_D}$ from Ref. \cite{Piotrowska:2017rgt} with $g_{V_D}\in[4.88,6.66]$ will be used to obtain the predictions
of the $V_E \to PV$ decays. After further considering $\mathcal{B}(K^*(1410)\to PV)\leq94.7\%$,
$\mathcal{B}(\rho(1450)\to PV)\leq100\%$ and $\mathcal{B}(\omega(1420)\to PV)\leq100\%$, the branching ratio predictions and the decay width predictions of the $V_E\to PV$ decays
 are obtained and  listed in Tab. \ref{Tab:VE2PVBr}. Our   predictions of the decay widths are consistent with ones in Ref. \cite{Piotrowska:2017rgt}, nevertheless, our errors of the predications are quite large, due to the  nonperturbative parameter $g_{V_E}=5.78\pm2.07$ and the decay widths of the excited vector mesons  such as  $\Gamma_{K^*(1410)}=232\pm21$ MeV,
$\Gamma_{\rho(1450)}=400\pm60$ MeV and $\Gamma_{\omega(1650)}=290\pm190$ MeV.
\begin{table}[t]
\renewcommand\arraystretch{0.8}
\tabcolsep 0.05in
\centering
\caption{Branching ratios and decay widths of the $V_E\to PV$ strong decays by the SU(3) flavor symmetry.}\vspace{0.08cm}
{\footnotesize
\begin{tabular}{lcc|lcc}  \hline\hline
$ V_E\to PV$ decays                        &  Branching ratios [\%]   & Decay width [MeV]                &Decay modes &  Branching ratios [\%]   & Decay width [MeV]\\\hline
$K^{*}(1410)^+\to \rho^0K^+$                &  $3.47\pm2.53$          & $7.58\pm5.25$                    &   $K^{*}(1410)^0\to \rho^-K^+$                &$6.94\pm5.06$       &  $15.16\pm10.51$       \\
$K^{*}(1410)^+\to \omega K^+$               &  $3.22\pm2.36$          & $7.03\pm4.89$                     &   $K^{*}(1410)^0\to \rho^0K^0$                &$3.35\pm2.45$         & $7.32\pm5.08$        \\
$K^{*}(1410)^+\to \rho^+ \overline{K}^0$    &  $6.70\pm4.90$          &  $14.64\pm10.17$                   &   $K^{*}(1410)^0\to \omega K^0$               &$3.10\pm2.27$         & $6.77\pm4.73$         \\
$K^{*}(1410)^+\to \phi K^+$                 &  $\cdots$               &  $\cdots$                          &   $K^{*}(1410)^0\to \phi K^0$                 & $\cdots$                &  $\cdots$             \\
$K^{*}(1410)^+\to K^{*+} \pi^0$             &  $8.06\pm5.61$          & $17.71\pm11.55$                    &   $K^{*}(1410)^0\to K^{*+} \pi^-$             &$16.00\pm11.14$       & $35.14\pm22.93$         \\
$K^{*}(1410)^+\to \overline{K}^{*0} \pi^+$  &  $15.69\pm10.94$        & $34.47\pm22.51$                    &   $K^{*}(1410)^0\to K^{*0} \pi^0$             &$7.91\pm5.51$       &  $17.37\pm11.34$        \\
$K^{*}(1410)^+\to K^{*+}\eta$               &  $\cdots$               & $\cdots$                           &   $K^{*}(1410)^0\to K^{*0}\eta$               &$\cdots$              &  $\cdots$           \\
$K^{*}(1410)^+\to K^{*+}\eta'$              &  $\cdots$               &  $\cdots$                          &   $K^{*}(1410)^0\to K^{*0}\eta'$              &$\cdots$              &  $\cdots$                        \\
$K^{*}(1410)\to K^{*}\pi$                   &  $23.83\pm16.60$        & $52.35\pm34.16$                   &   $K^{*}(1410)\to \omega K$                   &$3.16\pm2.31$         & $6.90\pm4.81$     \\
$K^{*}(1410)\to \rho K$                     &  $10.23\pm7.47$         & $22.35\pm15.50$                    &                                                                                               \\ \hline
$\rho(1450)^0\to \rho^0\eta$                 &$5.05\pm4.07$         & $19.56\pm15.22$                     &  $\rho(1450)^+\to \rho^+\eta$                  &$5.07\pm4.09$       &  $19.55\pm15.21$           \\
$\rho(1450)^0\to \rho^0\eta'$                &  $\cdots$               & $\cdots$                         &  $\rho(1450)^+\to \rho^+\eta'$                 &  $\cdots$          & $\cdots$                \\
$\rho(1450)^0\to \omega\pi^0$                &$32.93\pm22.86$        & $125.28\pm80.47$                   &  $\rho(1450)^+\to \omega\pi^+$                 &$32.78\pm22.77$     & $124.72\pm80.13$           \\
$\rho(1450)^0\to K^{*+}K^-$                  &$0.94\pm0.80$          & $3.43\pm2.78$                     &  $\rho(1450)^+\to K^{*+}\overline{K}^0$        &$1.78\pm1.52$       &   $6.48\pm5.30$           \\
$\rho(1450)^0\to K^{*-}K^+$                  &$0.94\pm0.80$          & $3.43\pm2.78$                     &  $\rho(1450)^+\to \overline{K}^{*0}K^+$        &$1.78\pm1.52$       &  $6.47\pm5.30$             \\
$\rho(1450)^0\to \overline{K}^{*0}K^0$       &$0.84\pm0.72$          & $3.05\pm2.52$                     &  $\rho(1450)\to \rho\eta$                      &$5.06\pm4.08$       & $19.56\pm15.21$              \\
$\rho(1450)^0\to K^{*0}\overline{K}^0$       &$0.84\pm0.72$          & $3.05\pm2.52$                     &   $\rho(1450)\to K^*K$                         &$3.55\pm3.03$       &  $12.95\pm10.60$             \\
$\rho(1450)\to \omega\pi$                    &$32.85\pm22.82$          & $125.00\pm80.30$                 &                                                                                                         \\ \hline
$\omega(1420)\to \omega\eta$                 &$2.17\pm2.06$          & $9.70\pm9.35$                    &  $\omega(1420)\to K^{*+}K^-$                  & $0.37\pm0.37$        & $1.65\pm1.65$            \\
$\omega(1420)\to \rho^0\eta'$                & $\cdots$               & $\cdots$                           &  $\omega(1420)\to K^{*-}K^+$                  & $0.37\pm0.37$        & $1.65\pm1.65$             \\
$\omega(1420)\to \rho^0\pi^0$                &$20.54\pm12.84$         & $95.67\pm60.95$                   &  $\omega(1420)\to \overline{K}^{*0}K^0$       & $0.32\pm0.32$        & $1.42\pm1.42$              \\
$\omega(1420)\to \rho^+\pi^-$                &$20.42\pm12.77$         & $95.14\pm60.57$                   &  $\omega(1420)\to K^{*0}\overline{K}^0$       & $0.32\pm0.32$       & $1.42\pm1.42$               \\
$\omega(1420)\to \rho^-\pi^+$                &$20.42\pm12.77$        & $95.14\pm60.57$                   &  $\omega(1420)\to K^*K$                       & $1.37\pm1.37$      & $6.14\pm6.14$               \\
$\omega(1420)\to \rho\pi$                    &$61.27\pm38.31$        & $285.42\pm181.72$                                                                                                                          \\\hline
\end{tabular}}\label{Tab:VE2PVBr}
\renewcommand\arraystretch{0.9}
\tabcolsep 0.06in
\centering
\caption{  Excited vector resonance state $V_{E/D}$ contributions to the branching ratios of the $B\to PV\ell^+\nu_{\ell}$ decays within 1$\sigma$ error  (in units of $10^{-4}$). }\vspace{0.08cm}
{\footnotesize
\begin{tabular}{l|ll|ll}  \hline\hline
Decay modes &  $\mathcal{B}_{[V_E]}$ with $\ell=\ell'$ & $\mathcal{B}_{[V_D]}$ with $\ell=\ell'$ &   $\mathcal{B}_{[V_E]}$ with $\ell=\tau$ & $\mathcal{B}_{[V_D]}$ with $\ell=\tau$ \\\hline
$B^+\to \rho^-\pi^+\ell^+\nu_{\ell}$                 &&&&  \\
or  $B^+\to \rho^+\pi^-\ell^+\nu_{\ell}$                              &  $0.26\pm0.18_{[\omega(1420)]}$       & $0.22\pm0.03_{[\omega(1650)]}$              & $0.12\pm0.08_{[\omega(1420)]}$        &$0.080\pm0.014_{[\omega(1650)]}$   \\
or $B^+\to \rho^0\pi^0\ell^+\nu_{\ell}$                &&&&\\\hline
$B^+\to \omega\eta\ell^+\nu_{\ell}$                    & $0.023\pm0.022_{[\omega(1420)]}$        & $0.057\pm0.015_{[\omega(1650)]}$              & $0.0098\pm0.0092_{[\omega(1420)]}$        & $0.021\pm0.005_{[\omega(1650)]}$                     \\\hline
%
%
$B^+\to \omega\pi^0\ell^+\nu_{\ell}$                   & $0.34\pm0.24_{[\rho(1450)]}$       & $0.26\pm0.09_{[\rho(1700)]}$               & $0.15\pm0.10_{[\rho(1450)]}$         & $0.094\pm0.031_{[\rho(1700)]}$                       \\\hline
$B^+\to \rho^0\eta\ell^+\nu_{\ell}$                    & $0.052\pm0.042_{[\rho(1450)]}$        & $0.081\pm0.032_{[\rho(1700)]}$              & $0.022\pm0.018_{[\rho(1450)]}$         & $0.029\pm0.011_{[\rho(1700)]}$                       \\\hline
$B^+\to \rho^0\eta'\ell^+\nu_{\ell}$                    & ~~~~~~~~~~~$\cdots$        & $0.00010\pm0.00010_{[\rho(1700)]}$              & ~~~~~~~~~~~$\cdots$         & $0.000036\pm0.000036_{[\rho(1700)]}$                       \\\hline
$B^+\to K^{*+}K^-\ell^+\nu_{\ell}$                     & $0.0097\pm0.0081_{[\rho(1450)]}$       &  $0.025\pm0.008_{[\rho(1700)]}$              & $0.0041\pm0.0034_{[\rho(1450)]}$        & $0.0090\pm0.0029_{[\rho(1700)]}$                        \\
or $B^+\to K^{*-}K^+\ell^+\nu_{\ell}$                                                       & $0.0039\pm0.0039_{[\omega(1420)]}$        & $0.018\pm0.003_{[\omega(1650)]}$              & $0.0017\pm0.0017_{[\omega(1420)]}$        & $0.0065\pm0.0011_{[\omega(1650)]}$                        \\\hline
$B^+\to K^{*0}\overline{K}^0\ell^+\nu_{\ell}$          & $0.0086\pm0.0074_{[\rho(1450)]}$       & $0.024\pm0.008_{[\rho(1700)]}$               & $0.0036\pm0.0031_{[\rho(1450)]}$        & $0.0087\pm0.0028_{[\rho(1700)]}$                           \\
or $B^+\to \overline{K}^{*0}K^0\ell^+\nu_{\ell}$                                                       & $0.0034\pm0.0034_{[\omega(1420)]}$          & $0.017\pm0.003_{[\omega(1650)]}$              & $0.0014\pm0.0014_{[\omega(1420)]}$        & $0.0063\pm0.0010_{[\omega(1650)]}$                           \\\hline
%
%
%
%
$B^0\to \omega\pi^-\ell^+\nu_{\ell}$                   &  $0.63\pm0.44_{[\rho(1450)]}$        &  $0.48\pm0.16_{[\rho(1700)]}$              &  $0.27\pm0.19_{[\rho(1450)]}$        &  $0.17\pm0.06_{[\rho(1700)]}$                             \\\hline
$B^0\to \rho^-\eta\ell^+\nu_{\ell}$                    &  $0.096\pm0.078_{[\rho(1450)]}$         &  $0.15\pm0.06_{[\rho(1700)]}$               &  $0.040\pm0.033_{[\rho(1450)]}$       &  $0.054\pm0.021_{[\rho(1700)]}$                            \\\hline
$B^0\to \rho^-\eta'\ell^+\nu_{\ell}$                    & ~~~~~~~~~~~$\cdots$          &  $0.00020\pm0.00020_{[\rho(1700)]}$               & ~~~~~~~~~~~$\cdots$         &  $0.000070\pm0.000070_{[\rho(1700)]}$                             \\\hline
$B^0\to K^{*0}K^-\ell^+\nu_{\ell}$                     &  $0.034\pm0.029_{[\rho(1450)]}$         &  $0.091\pm0.030_{[\rho(1700)]}$               &  $0.014\pm0.012_{[\rho(1450)]}$        &  $0.033\pm0.011_{[\rho(1700)]}$                             \\\hline
$B^0\to K^{*-}K^0\ell^+\nu_{\ell}$                     &  $0.034\pm0.029_{[\rho(1450)]}$         &  $0.092\pm0.030_{[\rho(1700)]}$              &  $0.014\pm0.012_{[\rho(1450)]}$        &  $0.033\pm0.011_{[\rho(1700)]}$                   \\\hline
$B_s^0\to K^{*-}\pi^0\ell^+\nu_{\ell}$                 &  $0.19\pm0.13_{[K^*(1410)]}$        &  $0.088\pm0.027_{[K^*(1680)]}$               &  $0.090\pm0.062_{[K^*(1410)]}$         &  $0.034\pm0.011_{[K^*(1680)]}$                  \\\hline
$B_s^0\to \overline{K}^{*0}\pi^-\ell^+\nu_{\ell}$      &  $0.38\pm0.26_{[K^*(1410)]}$        &  $0.17\pm0.05_{[K^*(1680)]}$              &  $0.18\pm0.12_{[K^*(1410)]}$         &  $0.067\pm0.021_{[K^*(1680)]}$                   \\\hline
$B_s^0\to \rho^0K^-\ell^+\nu_{\ell}$                   &  $0.082\pm0.058_{[K^*(1410)]}$        &  $0.069\pm0.021_{[K^*(1680)]}$              &  $0.038\pm0.027_{[K^*(1410)]}$         &  $0.026\pm0.008_{[K^*(1680)]}$                 \\\hline
$B_s^0\to \rho^-\overline{K}^0\ell^+\nu_{\ell}$        &  $0.16\pm0.11_{[K^*(1410)]}$        &  $0.14\pm0.04_{[K^*(1680)]}$              &  $0.073\pm0.052_{[K^*(1410)]}$         &  $0.052\pm0.016_{[K^*(1680)]}$                   \\\hline
$B_s^0\to \omega K^-\ell^+\nu_{\ell}$                  &  $0.076\pm0.054_{[K^*(1410)]}$        &  $0.067\pm0.020_{[K^*(1680)]}$              &  $0.035\pm0.025_{[K^*(1410)]}$         &  $0.026\pm0.008_{[K^*(1680)]}$                 \\\hline
$B_s^0\to K^{*-}\eta\ell^+\nu_{\ell}$                  & ~~~~~~~~~~~$\cdots$      &  $0.022\pm0.009_{[K^*(1680)]}$              & ~~~~~~~~~~~$\cdots$        &  $0.0082\pm0.0033_{[K^*(1680)]}$                 \\\hline
$B_s^0\to \phi K^-\ell^+\nu_{\ell}$                    & ~~~~~~~~~~~$\cdots$      &   $0.043\pm0.014_{[K^*(1680)]}$             & ~~~~~~~~~~~$\cdots$       &  $0.016\pm0.005_{[K^*(1680)]}$                 \\\hline
\end{tabular}}\label{Tab:B2PVlv1VED}
\end{table}

Now, in terms of the narrow width approximation given in Eq. (\ref{Eq:Br4BD}), $\mathcal{B}(B\to V_{E,D}\ell^+\nu_\ell)$ given in Tab. \ref{Tab:BrB2VEDlv}, and $\mathcal{B}(V_{E,D}\rightarrow PV)$ given in Tabs. \ref{Tab:VD2PVBr} and \ref{Tab:VE2PVBr},   the branching ratios of the $B\to V_{E,D}(V_{E,D}\to PV)\ell^+\nu_\ell$ are obtained, and  are listed in Tab. \ref{Tab:B2PVlv1VED}.
Comparing with the contributions of  the axial-vector resonance states and the tensor resonance states given in Tabs. \ref{Tab:B2PVlv1} and \ref{Tab:B2PVlv2},  one can see that only the $\rho(1700)$ resonance state contributes to the
$B^+\to \rho^0\eta'\ell^+\nu_{\ell}$    and  $B^0\to \rho^-\eta'\ell^+\nu_{\ell}$  decays, the branching ratios with some  excited vector resonance state  contributions  and  ones with the tensor resonance state contributions are  on the same order of magnitude;  other   excited vector resonance state  contributions  are at least 1 order larger than  the tensor ones.  In all $B\to PV\ell^+\nu_\ell$ decays except  $B\to \rho\eta'\ell^+\nu_{\ell}$, the dominant contributions still come from the  axial-vector resonance states.

Before we conclude,  two comments of the interference effects and the SU(3) flavor breaking effects are given.

Please note that the total branching ratios of the $B\to PV\ell^+\nu_{\ell}$ are not given in this work, since there are the interference terms between different resonant contributions and between non-resonant and resonant contributions, which we cannot determine by the SU(3) flavor symmetry now.  As discussed in Refs. \cite{Kang:2013jaa, Faller:2013dwa},  the interference effects  between different partial waves
vanish upon angular integration in the branching ratios.   Nevertheless,  there still are the interference effects between  different  resonances with the same partial wave. We take $B^+\to \rho^-\pi^+\ell^+\nu_{\ell}$ as an example, both $a_1(1260)^0$ and $h_1(1170)^0$ resonances give large contributions to $\mathcal{B}(B^+\to \rho^-\pi^+\ell^+\nu_{\ell})$, hence their interference terms  might be large.
Likewise,   the interference effects  might also be important for the $B^+\to \phi\eta\ell^+\nu_{\ell}$, $B\to K^{*}\overline{K}\ell^+\nu_{\ell}$, $B_s\to K^{*}\pi\ell^+\nu_{\ell}$, and $B_s^0\to K^{*-}\eta\ell^+\nu_{\ell}$   decays, in which the two or more kinds of contributions are important.

{ Although the interference effects  have not been studied in this work,  the total branching ratios  of the $B\to PV\ell^+\nu_\ell$ decays can be estimated based our results.
$\mathcal{B}(B^+\to \rho^\mp\pi^\pm\ell^+\nu_{\ell}/\rho^0\pi^0\ell'^+\nu_{\ell'}/\omega\pi^0\ell^+\nu_{\ell}/K^{*\mp}K^\pm\ell'^+\nu_{\ell'})$,  $\mathcal{B}(B^0\to \omega\pi^-\ell'^+\nu_{\ell'})$, and  $\mathcal{B}(B^0_s\to K^{*-}\pi^0\ell'^+\nu_{\ell'}/\overline{K}^{*0}\pi^-\ell^+\nu_{\ell}/\rho^-\overline{K}^0\ell'^+\nu_{\ell'})$ might  reach the order of $10^{-3}$.
$\mathcal{B}(B^+\to \rho^0\pi^0\tau^+\nu_{\tau}/\omega\eta\ell'^+\nu_{\ell'}/K^{*\mp}K^\pm\tau^+\nu_{\tau}/K^{*0}\overline{K}^0\ell^+\nu_\ell/\overline{K}^{*0}K^0\ell^+\nu_\ell)$, $\mathcal{B}(B^0\to \rho^0\pi^-\ell^+\nu_{\ell}/\rho^-\pi^0\ell^+\nu_{\ell}/\omega\pi^-\tau^+\nu_{\tau}/$ $\rho^-\eta\ell'^+\nu_{\ell'}/K^{*0}K^-\ell'^+\nu_{\ell'}/K^{*-}K^0\ell'^+\nu_{\ell'})$, and $\mathcal{B}(B^0_s\to \rho^0K^-\ell^+\nu_{\ell}/\omega K^-\ell^+\nu_{\ell}/K^{*-}\pi^0\tau^+\nu_{\tau}/\rho^-\overline{K}^0\tau^+\nu_{\tau})$ might  reach the order of $10^{-4}$.
$\mathcal{B}(B^+\to \rho^0\eta\ell^+\nu_{\ell}/\omega\eta\tau^+\nu_{\tau})$, $\mathcal{B}(B^0\to \rho^-\eta\tau^+\nu_{\tau}/K^{*0}K^-\tau^+\nu_{\tau}/K^{*-}K^0\tau^+\nu_{\tau})$, and $\mathcal{B}(B^0_s\to K^{*-}\eta\ell^+\nu_{\ell}/\phi K^-\ell'^+\nu_{\ell'})$     might  reach the order of $10^{-5}$.
$\mathcal{B}(B^+\to \phi\eta\ell^+\nu_{\ell})$  and $\mathcal{B}(B^0_s\to \phi K^-\tau^+\nu_{\tau})$  might  reach the order of $10^{-6}$.         }

In addition,  the SU(3) flavor breaking corrections due to a non-zero strange quark mass are not considered in this work. There are  very little experimental data in the  $B\to A/T/V_{E,D}\ell^+\nu_{\ell}$ and  $B\to PV\ell^+\nu_{\ell}$  decays, therefore, we cannot judge how large are the possible SU(3) breaking
effects. The SU(3) breaking effects are  usually of the order of
$\sim 20$\%  in some amplitudes, which  would conservatively expect an error
of 40\% in some branching ratios \cite{Xu:2013dta,He:2014xha,Yang:2015era,Carrillo-Serrano:2014zta,Geng:2018bow,Flores-Mendieta:1998tfv}.

\section{Summary}

The four-body semileptonic $B\to PV\ell^+\nu_\ell$  decays  are more difficult to calculate than the three-body semileptonic $B\to  M\ell^+\nu_\ell$  decays.
SU(3) flavor symmetry  can provide an opportunity
to relate different decay modes  and can  give some valuable information about the decays.
The charmless $B\to  PV\ell^+\nu_\ell$ decays with the axial-vector resonance states,   the tensor  resonance states,  and  the excited vector resonance states have been studied by the SU(3) flavor analysis in this work.
Our main results can be summarized as follows.

The  hadronic amplitudes of the $B \to A/T/V_{E,D} \ell^+\nu_{\ell}$ decays have been related by the nonperturbative parameters, and the branching ratio predictions of $B \to A/T/V_{E,D} \ell^+\nu_{\ell}$ have been obtained.
Most branching
ratios of the  $B \to A \ell^+\nu_{\ell}$ decays are   on the order of $\mathcal{O}(10^{-4}-10^{-3})$,  all branching
ratios of the  $B \to V_{E,D} \ell^+\nu_{\ell}$ decays are   on the order of $\mathcal{O}(10^{-5}-10^{-4})$,  most branching
ratios of the  $B \to T \ell^+\nu_{\ell}$ decays are   on the order of $\mathcal{O}(10^{-5})$,  and they are   reached
in current experiments.

In terms of the predictions of the $B \to A/T/V_{E,D} \ell^+\nu_{\ell}$  and $V_{E,D}\to PV$ decays in this work and the $A/T\to PV$ decays in Ref. \cite{Wang:2022fbk}, the branching ratios of the $B\to A(A\to PV)\ell^+\nu_\ell$,  $B\to T(T\to PV)\ell^+\nu_\ell$,  and $B\to V_{E,D}(V_{E,D}\to PV)\ell^+\nu_\ell$ decays have been obtained by   the narrow width approximation or further considering the width effects.   We have found that
$B^+\to \phi\eta\ell^+\nu_\ell$  only receive the contributions of the axial-vector meson resonances;
$B^+\to \rho^0\eta'\ell^+\nu_{\ell}$    and  $B^0\to \rho^-\eta'\ell^+\nu_{\ell}$  only receive the contributions of the excited vector resonances;
$B^+\to \rho^0\pi^0\ell^+\nu_\ell$, $B^+\to \omega\eta\ell^+\nu_\ell$, $B^+\to \omega\pi^0\ell^+\nu_\ell$, $B^+\to \rho^0\eta\ell^+\nu_\ell$,  $B^0\to \omega\pi^-\ell^+\nu_\ell$,
and $B^0\to \rho^-\eta\ell^+\nu_\ell$  receive the contributions of both axial-vector and excited vector meson resonances;  and other decays receive the contributions of all three kinds of    resonance contributions.
Some predicted branching ratios with the axial-vector meson resonances  are  on the order of $\mathcal{O}(10^{-4}-10^{-3})$, and other branching ratios with the axial-vector meson resonances and all branching ratios with the tensor and excited vector  meson resonances are small. After considering the resonant width effects of the axial-vector and tensor mesons, some branching ratios  are obviously  smaller than ones obtained by the  narrow width approximation.

SU(3) flavor  symmetry is a powerful tool in analyzing decays.
The branching ratios of the  $B\to A/T/V_{E,D}\ell^+\nu_{\ell}$  and $B\to A/T/V_{E,D}(A/T/V_{E,D}\to PV)\ell^+\nu_\ell$ decays have
been calculated with the SU(3) flavor symmetry. According to our rough predictions, some decays  could be tested in the future experiments such as  BelleII and LHCb.
They are
useful for testing and understanding the non-perturbative
effects in the semileptonic decays.

\section*{ACKNOWLEDGMENTS}
The work was supported by the
National Natural Science Foundation of China (Grants No. 12175088, No. 12305100, and
No. 12365014).

\section*{Appendix A: Hadronic amplitudes and form factors} \label{AppendixB}

The hadronic amplitudes  can be written as \cite{Sun:2011ssd,Li:2009tx,Colangelo:2019axi}
\begin{eqnarray}
H^A_{\pm} &=&(m_{B_q}-m_A)V^A_1(q^2)\mp\frac{2m_{B_q}|\vec{p}_A|}{(m_{B_q}-m_A)}A(q^2), \\
H^A_{0}&=& \frac{1}{2m_A\sqrt{q^2}}\left[(m_{B_q}^2-m_A^2-q^2)(m_{B_q}-m_A)V^A_1(q^2)-\frac{4m_{B_q}^2|\vec{p}_A|^2}{m_{B_q}-m_A}V^A_2(q^2)\right], \\
H^A_{t}&=& \frac{2m_{B_q}|\vec{p}_A|}{\sqrt{q^2}}V^A_0(q^2),
\end{eqnarray}
for the $B\to A\ell^+\nu_{\ell}$ decays,
\begin{eqnarray}
H^T_{\pm} &=&\frac{2|\vec{p}_T|}{2\sqrt{2}m_T}\left[(m_{B_q}+m_T)A^T_1(q^2)\mp\frac{2m_{B_q}|\vec{p}_T|}{(m_{B_q}+m_T)}V^T(q^2)\right], \\
H^T_{0}&=& \frac{|\vec{p}_T|}{\sqrt{2}m_T}\frac{1}{2m_T\sqrt{q^2}}\left[(m_{B_q}^2-m_T^2-q^2)(m_{B_q}+m_T)A^T_1(q^2)-\frac{4m_{B_q}^2|\vec{p}_T|^2}{m_{B_q}+m_T}A^T_2(q^2)\right], \\
H^T_{t}&=& \frac{|\vec{p}_T|}{\sqrt{2}m_T}\frac{2m_{B_q}|\vec{p}_T|}{\sqrt{q^2}}A^T_0(q^2),
\end{eqnarray}
for the $B\to T\ell^+\nu_{\ell}$ decays,
\begin{eqnarray}
H^{V'}_{\pm} &=&(m_{B_q}+m_{V'})A^{V'}_1(q^2)\mp\frac{2m_{B_q}|\vec{p}_{V'}|}{(m_{B_q}+m_{V'})}V(q^2), \\
H^{V'}_{0}&=& \frac{1}{2m_{V'}\sqrt{q^2}}\left[(m_{B_q}^2-m_{V'}^2-q^2)(m_{B_q}+m_{V'})A^{V'}_1(q^2)-\frac{4m_{B_q}^2|\vec{p}_{V'}|^2}{m_{B_q}+m_{V'}}A^{V'}_2(q^2)\right], \\
H^{V'}_{t}&=& \frac{2m_{B_q}|\vec{p}_{V'}|}{\sqrt{q^2}}A^{V'}_0(q^2),
\end{eqnarray}
for $B\to V'\ell^+\nu_\ell$ decays with $V'=V_E$ or $V_D$, where  $|\vec{p}_M|\equiv\sqrt{\lambda}/2m_{B_q}$.
The form factors   $A(q^2)$ and $V^A_{0,1,2,3}(q^2)$ ($V^T(q^2)$ and $A^T_{0,1,2}(q^2)$)  are given by  \cite{Cheng:2017pcq,Cheng:2003sm,Chen:2021ywv}
\begin{eqnarray}
\left<A(p,\varepsilon^*)\left|\bar{u}_k\gamma_{\mu}(1-\gamma_5)\bar{b}\right|B(p_B)\right>
&=&\frac{2iA(q^2)}{m_B-m_A}\epsilon_{\mu\nu\alpha\beta}\varepsilon^{*\nu}p^\alpha_Bp^\beta\nonumber\\
&&-\left[\varepsilon^*_\mu(m_B-m_A)V^A_1(q^2)-(p_B+p)_\mu(\varepsilon^*.p_B)\frac{V^A_2(q^2)}{m_B-m_A}\right]\nonumber\\
&&+q_\mu(\varepsilon^*.p_B)\frac{2m_A}{q^2}[V^A_3(q^2)-V^A_0(q^2)],\\
%
\left<T(p,\varepsilon^{*})\left|\bar{u}_k\gamma_{\mu}(1-\gamma_5)b\right|B(p_B)\right>
&=&\frac{2iV^T(q^2)}{m_B+m_T}\epsilon_{\mu\nu\alpha\beta}e^{*\nu}p^\alpha_Bp^\beta\nonumber\\
&&+2m_T\frac{e^*\cdot q}{q^2}q_\mu A^T_0(q^2)+(m_B+m_T)\Big(e^*_\mu-\frac{e^*\cdot q}{q^2}q_\mu\Big)A^T_1(q^2)\nonumber\\
&&-\frac{e^*\cdot q}{m_B+m_T}\Big((p_B+p)_\mu-\frac{m_B^2-m^2_T}{q^2}q_\mu\Big)A^T_2(q^2),\\
\left<V'(p,\varepsilon^*)\left|\bar{u}_k\gamma_{\mu}(1-\gamma_5)b\right|B(p_B)\right>
&=&\frac{2V^{V'}(q^2)}{m_B+m_{V'}}\epsilon_{\mu\nu\alpha\beta}\varepsilon^{*\nu}p^\alpha_Bp^\beta\nonumber\\
&&-i\left[\varepsilon^*_\mu(m_B+m_{V'})A^{V'}_1(q^2)-(p_B+p)_\mu(\varepsilon^*.p_B)\frac{A^{V'}_2(q^2)}{m_B+m_{V'}}\right]\nonumber\\
&&+iq_\mu(\varepsilon^*.p_B)\frac{2m_{V'}}{q^2}[A^{V'}_3(q^2)-A^{V'}_0(q^2)],
\end{eqnarray}
with  $e^{*\nu}\equiv \frac{\varepsilon^{*\mu\nu}\cdot p_{B\mu}}{m_B}$.

\begin{table}[h]
\renewcommand\arraystretch{1.2}
\tabcolsep 0.2in
\centering
\caption{Form factors of $B\to V_E$ transitions obtained by the light-front quark models \cite{CQPrepare},  $F(q^2)$ can be similarly obtained as the ones in Ref. \cite{Chen:2021ywv}, and $F^{B\to \omega(1420)}$ are the  same as $F^{B\to \rho(1450)}$. }
\begin{tabular}{lcc|lcc}  \hline\hline
Form factors                   &  $F(0)$      &  $b$  &   Form factors                   &  $F(0)$      &  $b$              \\\hline
$V^{B\to \rho(1450)}$          &  $0.33$    &$-1.07$  &   $V^{B_s\to K^*(1410)}$          &  $0.38$    &$-2.19$                 \\
$A_0^{B\to \rho(1450)}$        &  $0.28$    &$-1.25$  &   $A_0^{B_s\to K^*(1410)}$        &  $0.30$    &$-2.56$              \\
$A_1^{B\to \rho(1450)}$        &  $0.21$    &$5.43$   &   $A_1^{B_s\to K^*(1410)}$        &  $0.24$    &$3.18$               \\
$A_2^{B\to \rho(1450)}$        &  $0.16$    &$3.63$   &   $A_2^{B_s\to K^*(1410)}$        &  $0.20$    &$1.02$              \\     \hline
\end{tabular}\label{Tab:FF}
\end{table}

\section*{Appendix B: Meson multiplet} \label{AppendixA}

Bottom pseudoscalar triplet $B_i$ is composed of an anti-b quark and a light u/d/s quark, as follows:
\begin{eqnarray}
B_i=\Big( B^+(\bar{b}u),~B^0(\bar{b}d),~B_s^0(\bar{b}s)\Big).
\end{eqnarray}
The octet and the singlet components of axial-vector mesons are presented in the matrix of $A$ and $B$ \cite{Roca:2003uk},  as follows:
\begin{eqnarray}
A&=&\left(\begin{array}{ccc}
\frac{a^0_1}{\sqrt{2}}+\frac{f_1}{\sqrt{3}}+\frac{f_8}{\sqrt{6}} & a^+_1 & K^+_{1A} \\
a^-_1 &-\frac{a^0_1}{\sqrt{2}}+\frac{f_1}{\sqrt{3}}+\frac{f_8}{\sqrt{6}}  & K^0_{1A} \\
K^-_{1A} & \overline{K}^0_{1A} &\frac{f_1}{\sqrt{3}}-\frac{2f_8}{\sqrt{6}}
\end{array}\right)\,,\\
B&=&\left(\begin{array}{ccc}
\frac{b^0_1}{\sqrt{2}}+\frac{h_1}{\sqrt{3}}+\frac{h_8}{\sqrt{6}} & b^+_1 & K^+_{1B} \\
b^-_1 &-\frac{b^0_1}{\sqrt{2}}+\frac{h_1}{\sqrt{3}}+\frac{h_8}{\sqrt{6}}  & K^0_{1B} \\
K^-_{1B} & \overline{K}^0_{1B} &\frac{h_1}{\sqrt{3}}-\frac{2h_8}{\sqrt{6}}
\end{array}\right)\,.
\end{eqnarray}
where $A$ denotes the axial-vector mesons with the spin-parity quantum number of $J^{PC}=1^{++}$, and $B$ corresponds to the mesons with $J^{PC}=1^{+-}$.

Mesons $K_{1}(1270)$ and $K_{1}(1400)$ can be regarded as the mixtures of $K_{1A}$ and $K_{1B}$ with the mixing angle $\theta_{K_1}$,  as follows:

\begin{eqnarray}
\left(\begin{array}{c}
K_1(1270)\\
K_1(1400)
\end{array}\right)\,
=
\left(\begin{array}{cc}
sin\theta_{K_1}&cos\theta_{K_1}\\
cos\theta_{K_1}&-sin\theta_{K_1}
\end{array}\right)\,
\left(\begin{array}{c}
K_{1A}\\
K_{1B}
\end{array}\right).
\end{eqnarray}
The value range of $\theta_{K_1}$ has been investigated in our previous work; $~\theta_{K_1}\in[52^\circ, 65^\circ]$ within 1$\sigma$ error \cite{Wang:2022fbk} will be used in this work.

Mesons  $f_1(1285)$, $f_1(1420)$ can be expressed by the mixing angle  $\theta_{3P_1}$ while $h_1(1170)$, $h_1(1415)$ are mixed by $\theta_{1P_1}$, as follows:
\begin{eqnarray}
\left(\begin{array}{c}
f_1(1285)\\
f_1(1420)
\end{array}\right)\,
=
\left(\begin{array}{cc}
cos\theta_{3P_1}&sin\theta_{3P_1}\\
-sin\theta_{3P_1}&cos\theta_{3P_1}
\end{array}\right)\,
\left(\begin{array}{c}
f_{1}\\
f_{8}
\end{array}\right),\label{Eq:Mixf1}\\
\left(\begin{array}{c}
h_1(1170)\\
h_1(1415)
\end{array}\right)\,
=
\left(\begin{array}{cc}
cos\theta_{1P_1}&sin\theta_{1P_1}\\
-sin\theta_{1P_1}&cos\theta_{1P_1}
\end{array}\right)\,
\left(\begin{array}{c}
h_{1}\\
h_{8}
\end{array}\right).\label{Eq:Mixh1}
\end{eqnarray}
where $~\theta_{1P_1}\in[5^\circ, 56^\circ],~ \theta_{3P_1}\in[56^\circ, 125^\circ]$ within 1$\sigma$ error \cite{Wang:2022fbk} will be used in our analysis.
There are different definitions of the mixing angles for $f_1$ and $h_1$ \cite{Kang:2018jzg}
\begin{eqnarray}
A_L=sin\alpha_AA_q+cos\alpha_AA_s,\label{Eq:alpha1}\\
A_H=cos\alpha_AA_q-sin\alpha_AA_s,\label{Eq:alpha2}
\end{eqnarray}
where $A_L(A_H)$ is the light(heavier) axial vector, $A_q$ and  $A_s$ denotes the corresponding components $(u\bar{u}+d\bar{d})/\sqrt{2}$ and $s\bar{s}$  in the wave functions.
 $\alpha_{f_1}=(69.7\pm8)^\circ$ and $\alpha_{h_1}=(86.7\pm6)^\circ$  are also analyzed in this work \cite{Kang:2018jzg}.

Light pseudoscalar mesons and vector mesons  are represented as
 \cite{He:2018joe}
\begin{eqnarray}
 P&=&\left(\begin{array}{ccc}
\frac{\pi^0}{\sqrt{2}}+\frac{\eta_8}{\sqrt{6}}+\frac{\eta_1}{\sqrt{3}} & \pi^+ & K^+ \\
\pi^- &-\frac{\pi^0}{\sqrt{2}}+\frac{\eta_8}{\sqrt{6}}+\frac{\eta_1}{\sqrt{3}}  & K^0 \\
K^- & \overline{K}^0 &-\frac{2\eta_8}{\sqrt{6}}+\frac{\eta_1}{\sqrt{3}}
\end{array}\right)\,,\\
V&=&\left(\begin{array}{ccc}
\frac{\rho^0}{\sqrt{2}}+\frac{\omega}{\sqrt{2}} & \rho^+ & K^{*+} \\
\rho^- &-\frac{\rho^0}{\sqrt{2}}+\frac{\omega}{\sqrt{2}} & K^{*0} \\
K^{*-} & \overline{K}^{*0} &\phi
\end{array}\right)\,.\label{Eq:VM}
\end{eqnarray}
The excited vector mesons $V_E$ and $V_D$ are similar to $V$ given in Eq. (\ref{Eq:VM}).
 The pseudoscalar mesons $\eta$ and $\eta'$ are the mixtures of $\eta_1=\frac{u\bar{u}+d\bar{d}+s\bar{s}}{\sqrt{3}}$ and $\eta_8=\frac{u\bar{u}+d\bar{d}-2s\bar{s}}{\sqrt{6}}$ with a mixing angle denoted as $\theta_P$ \cite{Bramon:1992kr}, as follows:
\begin{eqnarray}
\left(\begin{array}{c}
\eta\\
\eta'
\end{array}\right)\,
=
\left(\begin{array}{cc}
\mbox{cos}\theta_P&-\mbox{sin}\theta_P\\
\mbox{sin}\theta_P&\mbox{cos}\theta_P
\end{array}\right)\,\left(\begin{array}{c}
\eta_8\\
\eta_1
\end{array}\right)\,.
\end{eqnarray}
In our numerical analysis, we will utilize the mixing angle range of $\theta_P=[-20^\circ,-10^\circ]$ from the PDG \cite{PDG2024}.

The exchange of the lowest multiplet of p-wave tensor mesons with $J^{PC}=2^{++}$ takes the form   \cite{Ecker:2007us,Chen:2023ybr}
\begin{eqnarray}
 T&=&\left(\begin{array}{ccc}
\frac{a_2^0}{\sqrt{2}}+\frac{f_2^q}{\sqrt{2}} & a^+_2 & K_2^{*+}  \\
a^-_2 & -\frac{a^0_2}{\sqrt{2}}+\frac{f^q_2}{\sqrt{2}} & K^{*0}_2 \\
K^{*-}_2 & \overline{K}^{*0}_2 & f^s_2
\end{array}\right)\,.
\end{eqnarray}
In this paper, the isovector mesons $a_2(1320)$, isodoublet states $K_2^*(1430)$ and two isosinglet mesons $f_2(1270)$ and  $f_2'(1525)$ are investigated; $f_{2}(1270)$ and $f_2'(1525)$ are mixed by $f_2^q$ and $f_2^s$ with the mixing angle $\theta_{f_2}$, as follows:
\begin{eqnarray}
\left(\begin{array}{c}
f_2(1270)\\
f_2'(1525)
\end{array}\right)\,
=
\left(\begin{array}{cc}
cos\theta_{f_2}&sin\theta_{f_2}\\
sin\theta_{f_2}&-cos\theta_{f_2}
\end{array}\right)\,
\left(\begin{array}{c}
f^q_2\\
f^s_2
\end{array}\right),
\end{eqnarray}
where $~\theta_{f_2}\in[8^\circ, 10^\circ]$ \cite{Cheng:2010yd} will be used in our calculation.

\section*{References}


\begin{thebibliography}{99}



\bibitem{He:1998rq}
  X.~G.~He,
  Eur.\ Phys.\ J.\ C {\bf 9}, 443 (1999)
  [hep-ph/9810397].

\bibitem{He:2000ys}
  X.~G.~He, Y.~K.~Hsiao, J.~Q.~Shi, Y.~L.~Wu and Y.~F.~Zhou,
  Phys.\ Rev.\ D {\bf 64}, 034002 (2001)
  [hep-ph/0011337].


\bibitem{Fu:2003fy}
  H.~K.~Fu, X.~G.~He and Y.~K.~Hsiao,
  Phys.\ Rev.\ D {\bf 69}, 074002 (2004)
  [hep-ph/0304242].

\bibitem{Hsiao:2015iiu}
  Y.~K.~Hsiao, C.~F.~Chang and X.~G.~He,
  Phys.\ Rev.\ D {\bf 93}, 114002 (2016)
  [arXiv:1512.09223 [hep-ph]].

\bibitem{He:2015fwa}
  X.~G.~He and G.~N.~Li,
  Phys.\ Lett.\ B {\bf 750}, 82 (2015)
  [arXiv:1501.00646 [hep-ph]].

\bibitem{Gronau:1994rj}
  M.~Gronau, O.~F.~Hernandez, D.~London and J.~L.~Rosner,
  Phys.\ Rev.\ D {\bf 50}, 4529 (1994)
  [hep-ph/9404283].

\bibitem{Gronau:1995hm}
  M.~Gronau, O.~F.~Hernandez, D.~London and J.~L.~Rosner,
  Phys.\ Rev.\ D {\bf 52}, 6356 (1995)
  [hep-ph/9504326].

\bibitem{Zhou:2016jkv}
  S.~H.~Zhou, Q.~A.~Zhang, W.~R.~Lyu and C.~D.~L,
  Eur.\ Phys.\ J.\ C {\bf 77}, 125 (2017)
  [arXiv:1608.02819 [hep-ph]].


\bibitem{Cheng:2014rfa}
  H.~Y.~Cheng, C.~W.~Chiang and A.~L.~Kuo,
  Phys.\ Rev.\ D {\bf 91}, 014011 (2015)
  [arXiv:1409.5026 [hep-ph]].



\bibitem{He:2015fsa}
  M.~He, X.~G.~He and G.~N.~Li,
  Phys.\ Rev.\ D {\bf 92}, 036010 (2015)
  [arXiv:1507.07990 [hep-ph]].

 \bibitem{Deshpande:1994ii}
  N.~G.~Deshpande and X.~G.~He,
  Phys.\ Rev.\ Lett.\  {\bf 75}, 1703 (1995)
  [hep-ph/9412393].

  \bibitem{Shivashankara:2015cta}
  S.~Shivashankara, W.~Wu and A.~Datta,
  Phys.\ Rev.\ D {\bf 91},  115003 (2015)
  [arXiv:1502.07230 [hep-ph]].

\bibitem{Wang:2021uzi}
R.~M.~Wang, Y.~G.~Xu, C.~Hua and X.~D.~Cheng,
Phys. Rev. D \textbf{103}, 013007  (2021)
[arXiv:2101.02421 [hep-ph]].

\bibitem{Wang:2020wxn}
R.~M.~Wang, X.~D.~Cheng, Y.~Y.~Fan, J.~L.~Zhang and Y.~G.~Xu,
J. Phys. G \textbf{48}, 085001  (2021)
[arXiv:2008.06624 [hep-ph]].

  \bibitem{Grossman:2012ry}
  Y.~Grossman and D.~J.~Robinson,
  JHEP {\bf 1304}, 067 (2013)
  [arXiv:1211.3361 [hep-ph]].


  \bibitem{Pirtskhalava:2011va}
  D.~Pirtskhalava and P.~Uttayarat,
  Phys.\ Lett.\ B {\bf 712}, 81 (2012)
  [arXiv:1112.5451 [hep-ph]].


\bibitem{Cheng:2012xb}
  H.~Y.~Cheng and C.~W.~Chiang,
  Phys.\ Rev.\ D {\bf 86}, 014014 (2012)
  [arXiv:1205.0580 [hep-ph]].

\bibitem{Savage:1989qr}
  M.~J.~Savage and R.~P.~Springer,
  Phys.\ Rev.\ D {\bf 42}, 1527 (1990).


\bibitem{Savage:1991wu}
  M.~J.~Savage,
  Phys.\ Lett.\ B {\bf 257}, 414 (1991).



\bibitem{Altarelli:1975ye}
 G. Altarelli, N. Cabibbo, and L. Maiani, Phys. Lett. B {\bf 57}, 277 (1975).

\bibitem{Lu:2016ogy}
  C.~D.~L\"{u}, W.~Wang and F.~S.~Yu,
  Phys.\ Rev.\ D {\bf 93},   056008 (2016)
  [arXiv:1601.04241 [hep-ph]].


\bibitem{Geng:2017esc}
  C.~Q.~Geng, Y.~K.~Hsiao, Y.~H.~Lin and L.~L.~Liu,
  Phys.\ Lett.\ B {\bf 776}, 265 (2018)
  [arXiv:1708.02460 [hep-ph]].

\bibitem{Geng:2018plk}
  C.~Q.~Geng, Y.~K.~Hsiao, C.~W.~Liu and T.~H.~Tsai,
  Phys.\ Rev.\ D {\bf 97},   073006 (2018)
  [arXiv:1801.03276 [hep-ph]].

\bibitem{Geng:2017mxn}
  C.~Q.~Geng, Y.~K.~Hsiao, C.~W.~Liu and T.~H.~Tsai,
  JHEP {\bf 1711}, 147 (2017)
  [arXiv:1709.00808 [hep-ph]].

\bibitem{Geng:2019bfz}
  C.~Q.~Geng, C.~W.~Liu, T.~H.~Tsai and S.~W.~Yeh,   Phys. Lett.
B {\bf 792}, 214 (2019)
  [arXiv:1901.05610 [hep-ph]].


\bibitem{Wang:2017azm}
  W.~Wang, Z.~P.~Xing and J.~Xu,
  Eur.\ Phys.\ J.\ C {\bf 77},   800 (2017)
  [arXiv:1707.06570 [hep-ph]].

\bibitem{Wang:2019dls}
D.~Wang,
Eur. Phys. J. C \textbf{79},   429 (2019)
[arXiv:1901.01776 [hep-ph]].


\bibitem{Wang:2017gxe}
  D.~Wang, P.~F.~Guo, W.~H.~Long and F.~S.~Yu,
  JHEP {\bf 1803}, 066 (2018)
  [arXiv:1709.09873 [hep-ph]].


\bibitem{Muller:2015lua}
  S.~M\"{u}ller, U.~Nierste and S.~Schacht,
  Phys.\ Rev.\ D {\bf 92},   014004 (2015)
  [arXiv:1503.06759 [hep-ph]].






\bibitem{Qiao:2024nbq}
Y.~Qiao, Y.~X.~Liu, Y.~G.~Xu and R.~M.~Wang,
Eur. Phys. J. C \textbf{84},   1110 (2024)
[arXiv:2404.03857 [hep-ph]].


\bibitem{Wang:2022fbk}
R.~M.~Wang, Y.~Qiao, Y.~J.~Zhang, X.~D.~Cheng and Y.~G.~Xu,
Phys. Rev. D 107, 056022 (2023)
[arXiv:2301.00090 [hep-ph]].

\bibitem{Feldmann:2018kqr}
T.~Feldmann, D.~Van Dyk and K.~K.~Vos,
JHEP \textbf{10}, 030 (2018)
[arXiv:1807.01924 [hep-ph]].

\bibitem{Faller:2013dwa}
S.~Faller, T.~Feldmann, A.~Khodjamirian, T.~Mannel and D.~van Dyk,
Phys. Rev. D \textbf{89},  014015 (2014)
[arXiv:1310.6660 [hep-ph]].


\bibitem{Kim:2017dfr}
C.~S.~Kim, G.~L.~Castro and S.~L.~Tostado,
Phys. Rev. D \textbf{95},   073003 (2017)
[arXiv:1702.01704 [hep-ph]].


\bibitem{Ananthanarayan:2005us}
B.~Ananthanarayan and K.~Shivaraj,
Phys. Lett. B \textbf{628}, 223  (2005)
[arXiv:hep-ph/0508116 [hep-ph]].





\bibitem{Ivanov:2019nqd}
M.~A.~Ivanov, J.~G.~K\"orner, J.~N.~Pandya, P.~Santorelli, N.~R.~Soni and C.~T.~Tran,
Front. Phys. (Beijing) \textbf{14}, 64401 (2019)
[arXiv:1904.07740 [hep-ph]].




\bibitem{Xu:2013dta}
D.~Xu, G.~N.~Li and X.~G.~He,
Int. J. Mod. Phys. A \textbf{29}, 1450011  (2014)
[arXiv:1307.7186 [hep-ph]].


\bibitem{He:2014xha}
X.~G.~He, G.~N.~Li and D.~Xu,
Phys. Rev. D \textbf{91}, 014029  (2015)
[arXiv:1410.0476 [hep-ph]].


\bibitem{PDG2024}
S. Navas et al. (Particle Data Group), Phys. Rev. D {\bf 110}, 030001 (2024)

\bibitem{Verma:2011yw}
R.~C.~Verma,
J. Phys. G \textbf{39}, 025005 (2012)
[arXiv:1103.2973 [hep-ph]].





\bibitem{Kang:2018jzg}
X.~W.~Kang, T.~Luo, Y.~Zhang, L.~Y.~Dai and C.~Wang,
Eur. Phys. J. C \textbf{78},  909 (2018)
[arXiv:1808.02432 [hep-ph]].



\bibitem{Chen:2021ywv}
L.~Chen, Y.~W.~Ren, L.~T.~Wang and Q.~Chang,
Eur. Phys. J. C \textbf{82} (2022), 451
[arXiv:2112.08016 [hep-ph]].



\bibitem{Hatanaka:2010fpr}
H.~Hatanaka and K.~C.~Yang,
Eur. Phys. J. C \textbf{67}, 149  (2010)
[arXiv:0907.1496 [hep-ph]].

\bibitem{CQPrepare}
Liting Wang, Qin Chang, in preparing.


\bibitem{Cheng:1993ah}
H.~Y.~Cheng, C.~Y.~Cheung, W.~Dimm, G.~L.~Lin, Y.~C.~Lin, T.~M.~Yan and H.~L.~Yu,
Phys. Rev. D \textbf{48}, 3204  (1993)
[arXiv:hep-ph/9305340 [hep-ph]].

\bibitem{Tsai:2021ota}
S.~Y.~Tsai and Y.~K.~Hsiao,
[arXiv:2107.03634 [hep-ph]].

\bibitem{Liang:2024lon}
J.~Liang, S.~Chen, Y.~Chen, C.~Shi and W.~Sun,
Sci. China Phys. Mech. Astron. \textbf{68},  251011 (2025)
[arXiv:2409.14410 [hep-lat]].

\bibitem{Piotrowska:2017rgt}
M.~Piotrowska, C.~Reisinger and F.~Giacosa,
Phys. Rev. D \textbf{96}, 054033 (2017)
[arXiv:1708.02593 [hep-ph]].


\bibitem{Kang:2013jaa}
X.~W.~Kang, B.~Kubis, C.~Hanhart and U.~G.~Mei\ss{}ner,
Phys. Rev. D \textbf{89}, 053015 (2014)
[arXiv:1312.1193 [hep-ph]].

\bibitem{Yang:2015era}
G.~S.~Yang and H.~C.~Kim,
Phys. Rev. C \textbf{92}, 035206 (2015)
[arXiv:1504.04453 [hep-ph]].

\bibitem{Carrillo-Serrano:2014zta}
M.~E.~Carrillo-Serrano, I.~C.~Clo\"et and A.~W.~Thomas,
Phys. Rev. C \textbf{90},  064316 (2014)
[arXiv:1409.1653 [nucl-th]].


\bibitem{Geng:2018bow}
C.~Q.~Geng, Y.~K.~Hsiao, C.~W.~Liu and T.~H.~Tsai,
Eur. Phys. J. C \textbf{78}, 593 (2018)
[arXiv:1804.01666 [hep-ph]].


\bibitem{Flores-Mendieta:1998tfv}
R.~Flores-Mendieta, E.~E.~Jenkins and A.~V.~Manohar,
Phys. Rev. D \textbf{58}, 094028 (1998)
[arXiv:hep-ph/9805416 [hep-ph]].




\bibitem{Sun:2011ssd}
Y.~J.~Sun, Z.~G.~Wang and T.~Huang,
Chin. Phys. C \textbf{36}, 1046 (2012)
[arXiv:1106.4915 [hep-ph]].

\bibitem{Li:2009tx}
R.~H.~Li, C.~D.~Lu and W.~Wang,
Phys. Rev. D \textbf{79}, 034014 (2009)
[arXiv:0901.0307 [hep-ph]].


\bibitem{Colangelo:2019axi}
P.~Colangelo, F.~De Fazio and F.~Loparco,
Phys. Rev. D \textbf{100}, 075037 (2019)
[arXiv:1906.07068 [hep-ph]].

\bibitem{Cheng:2017pcq}
H.~Y.~Cheng and X.~W.~Kang,
Eur. Phys. J. C \textbf{77}, 587  (2017)
[erratum: Eur. Phys. J. C \textbf{77}, 863  (2017)]
[arXiv:1707.02851 [hep-ph]].



\bibitem{Cheng:2003sm}
H.~Y.~Cheng, C.~K.~Chua and C.~W.~Hwang,
Phys. Rev. D \textbf{69}, 074025  (2004)
[arXiv:hep-ph/0310359 [hep-ph]].



\bibitem{Roca:2003uk}
L.~Roca, J.~E.~Palomar and E.~Oset,
Phys. Rev. D \textbf{70}, 094006  (2004)
[arXiv:hep-ph/0306188 [hep-ph]].


\bibitem{He:2018joe}
X.~G.~He, Y.~J.~Shi and W.~Wang,
Eur. Phys. J. C \textbf{80},   359 (2020)
[arXiv:1811.03480 [hep-ph]].


\bibitem{Bramon:1992kr}
A.~Bramon, A.~Grau and G.~Pancheri,
Phys. Lett. B \textbf{283}, 416   (1992).

\bibitem{Chen:2023ybr}
C.~Chen, N.~Q.~Cheng, L.~W.~Yan, C.~G.~Duan and Z.~H.~Guo,
Phys. Rev. D \textbf{108}, 014002  (2023)
[arXiv:2302.11316 [hep-ph]].



\bibitem{Ecker:2007us}
G.~Ecker and C.~Zauner,
Eur. Phys. J. C \textbf{52}, 315   (2007)
[arXiv:0705.0624 [hep-ph]].


\bibitem{Cheng:2010yd}
H.~Y.~Cheng and K.~C.~Yang,
Phys. Rev. D \textbf{83}, 034001  (2011)
[arXiv:1010.3309 [hep-ph]].






\end{thebibliography}
\end{document}